\documentclass[aps,prb,onecolumn,floatfix]{revtex4}

\usepackage{graphicx}
\usepackage{amssymb,amsfonts,amsmath}
\usepackage{epsf}
\usepackage{subfigure}
\usepackage{epstopdf}

\newcommand{\be}{\begin{equation}}
\newcommand{\ee}{\end{equation}}
\newcommand{\bea}{\begin{eqnarray}}
\newcommand{\eea}{\end{eqnarray}}

\newcommand{\paren}[1]{\left( #1 \right)}

\begin{document}

\title{\bf Kinetics and thermodynamics of first-order Markov chain copolymerization}

\author{P. Gaspard and D. Andrieux\footnote{Presently at Sopra Banking Software.}}
\affiliation{Center for Nonlinear Phenomena and Complex Systems,\\
Universit\'e Libre de Bruxelles, Code Postal 231, Campus Plaine,
B-1050 Brussels, Belgium}

\begin{abstract}
We report a theoretical study of stochastic processes modeling the growth of first-order
Markov copolymers, as well as the reversed reaction of depolymerization.
These processes are ruled by kinetic equations describing both the attachment
and detachment of monomers. Exact solutions are obtained for these kinetic equations in the 
steady regimes of multicomponent copolymerization and depolymerization. 
Thermodynamic equilibrium is identified as the state at which the growth velocity
is vanishing on average and where detailed balance is satisfied. 
Away from equilibrium, the analytical expression of the thermodynamic entropy production is deduced in terms of the Shannon disorder per monomer in the copolymer sequence.  The Mayo-Lewis equation
is recovered in the fully irreversible growth regime.  The theory also applies
to Bernoullian chains in the case where the attachment and detachment rates 
only depend on the reacting monomer.
\end{abstract}

\maketitle

\section{Introduction}

Copolymerization processes are most often described in the fully irreversible growth regime
since pioneering works in the forties.\cite{ML44,AG44,FR99}  These approaches consider
the attachment of monomers to the end of the growing copolymer but neglect the possible
detachment.  Such an approximation is no longer valid close to the so-called ceiling temperature where the detachment rates become comparable to the attachment rates.\cite{DI48}
At this critical temperature, the growth velocity vanishes and the process is at equilibrium
because detailed balancing is then satisfied between the attachment and detachment of monomers.
However, the kinetic equations are much more difficult to solve if the detachment rates
are not negligible.

In this paper, we show that these kinetic equations can nevertheless
be solved analytically in the steady regimes of living copolymerization or depolymerization
under the assumption that the attachment and detachment rates depend on the last monomeric unit
at the tip of the reactive copolymer.  The remarkable result is that the growing copolymer
is exactly described as a first-order Markov chain, the properties of which can be deduced from the
kinetic equations.  These exact solutions allow us to obtain the
analytical expression for the thermodynamic entropy production in terms of the mean growth velocity
and the associated affinity.  This latter is the sum of the free-energy driving force and the
Shannon disorder per monomer of the first-order Markov chain describing the copolymer sequence.\cite{AG08,AG09}  In the fully irreversible regime where the detachment rates are negligible, the
Mayo-Lewis equation is recovered for the ratio of mole fractions in the copolymer.\cite{ML44,AG44,FR99}
Furthermore, in the case where the attachment and detachment rates do not depend on the previous monomeric units, the growing chain is Bernoullian and our previous results are also recovered.\cite{AG09}

The paper is organized as follows.  In Section~\ref{Equations}, the kinetic equations ruling the process are introduced.  These equations are solved in the steady regime of copolymerization in Section~\ref{Copolym} where the expression for the thermodynamic entropy production is obtained.  The state of thermodynamic equilibrium is identified in Section~\ref{Thermo}.  The regime of depolymerization and its thermodynamics are presented in Section~\ref{Depolym}.  Examples are given in Section~\ref{Examples} where the analytical results are compared with numerical simulations.  Conclusions are drawn in Section~\ref{Conclusions}.  The reduction to the case of Bernoullian chains is carried out in Appendix~\ref{AppA}.

\section{The kinetic equations}
\label{Equations}

We consider a process of living copolymerization, i.e., copolymerization without termination.
The process takes place in a solution where the copolymers are sufficiently diluted to be independent of each other.  Accordingly, the process may be described at the level of a single copolymer.  We adopt a coarse-graining description in terms of the sequence $\omega$ of monomeric units $m_i=1,2,...,M$ composing the copolymer at a given time $t$.  $M$ denotes the number of different types of monomers.  The possible sequences encountered during the process are given by
\be
\omega\in\{\emptyset, \, m_1, \, m_1m_2, \, m_1m_2m_3,\, ...,\, m_1m_2\cdots m_l,\, ...\} \, .
\label{seq}
\ee
The length of the copolymer is measured as the number $l=\vert\omega\vert$ of monomeric units.

This copolymer evolves in time due to the random events of attachment and detachment of monomers:
\be
\omega=m_1m_2 \cdots m_{l-1} \ + \ m_l 
\quad \underset{w_{-m_l\vert m_{l-1}}}{\overset{w_{+m_l\vert m_{l-1}}}{\rightleftharpoons}} \quad 
\omega'=m_1m_2 \cdots m_{l-1}m_l \, .
\ee
The rates of attachment and detachment, $w_{\pm m_l\vert m_{l-1}}$, are assumed to depend on the last
monomeric unit $m_{l-1}$ at the tip of the copolymer. The rates obey the mass action law.  Accordingly, every attachment rate $w_{+m_l\vert m_{l-1}}$ is proportional to the concentration $[m_l]$ of the monomer that attaches, while the detachment rates $w_{-m_l\vert m_{l-1}}$ are independent of the concentrations:
\begin{subequations}
\label{mass_action_rates}
\bea
w_{+m_l\vert m_{l-1}} &=& k_{+m_l\vert m_{l-1}} \ [m_l] \, ,\\
w_{-m_l\vert m_{l-1}} &=& k_{-m_l\vert m_{l-1}} \, ,
\eea
\end{subequations}
with the rate constants $k_{\pm m_l\vert m_{l-1}}$.  The solution surrounding the copolymer is supposed to constitute a large enough pool of monomers so that their concentrations $[m]$ remain constant during the whole process.

The initiation of the copolymerization process is determined by the rates $w_{\pm m_1\vert \emptyset}$  of transition between $\omega=\emptyset$ and $\omega'=m_1$.

Since the process is stochastic for a single copolymer, its time evolution is described in terms of the probability $P_t(m_1m_2\cdots m_l)$ to find the copolymer with the sequence $\omega=m_1m_2\cdots m_l$ at the time $t$.  These probabilities are ruled by the kinetic equations
\bea
\frac{d}{dt}\, P_t(m_1 \cdots m_{l-1}m_l) &=& w_{+m_l\vert m_{l-1}} \, P_t(m_1\cdots m_{l-1}) 
+\sum_{m_{l+1}=1}^M w_{-m_{l+1}\vert m_{l}} \, P_t(m_1\cdots m_{l-1}m_{l}m_{l+1}) \nonumber\\
&-& \left( w_{-m_{l}\vert m_{l-1}} + \sum_{m_{l+1}=1}^M w_{+m_{l+1}\vert m_{l}}\right)  P_t(m_1\cdots m_{l-1}m_{l}) \, ,
\label{kinetic_eqs}
\eea
which form an infinite hierarchy of coupled ordinary differential equations.  During copolymerization, the probability mass moves to longer and longer copolymer sequences.  The total probability is conserved by the kinetic equations for all time, $\sum_{\omega} P_t(\omega) = 1$, where the sum extends over the sequences (\ref{seq}).  Indeed, the kinetic equations (\ref{kinetic_eqs}) satisfy this normalization condition given that $w_{\pm\emptyset\vert\emptyset}=0$.

In a dilute solution, the concentrations of the different sequences of copolymers obey the same kinetic equations because the concentrations are proportional to the probabilities if the copolymers are independent of each other.  Indeed, for a solution containing $N$ independent copolymers in a volume $V$, the concentrations are related to the probabilities by
\be
[m_1m_2\cdots m_l ]_t = \frac{N}{V} \, P_t(m_1 \cdots m_l) \, .
\ee

The thermodynamics of general copolymerization processes was studied in Refs.~\onlinecite{AG08} and~\onlinecite{AG09}, where the expression of entropy production was obtained that will be used below for the processes we here consider.

In the next section, the kinetic equations~(\ref{kinetic_eqs}) are solved in the regime of copolymerization.  Section~\ref{Thermo} is devoted to the critical situation of thermodynamic equilibrium and the regime of depolymerization is treated in Section~\ref{Depolym}.

\section{Copolymerization}
\label{Copolym}

\subsection{Solving the kinetic equations}

The growth process of the length $l$ of the copolymer is similar to a random walk with a drift controlled by the changing last monomeric unit of the chain.  The front of this growth process propagates at a mean velocity $v$ so that the length $l$ is distributed around the average length $\langle l\rangle_t \simeq v \, t$.  The variance of the length is expected to increase linearly in time as $\langle l^2\rangle_t -\langle l\rangle_t^2 \simeq 2\, {\cal D} \, t$ with the diffusivity $\cal D$.  The sequence of the copolymer has statistical properties that can be supposed to be stationary after a long enough time.  In this regime, we may assume that the solution of the kinetic equations takes the form
\be
P_t(m_1\cdots m_{l-1}m_{l}) \simeq \mu(m_1\cdots m_{l-1}m_{l}) \, p_t(l) \, ,
\label{Hyp}
\ee
where
\be
p_t(l) \equiv \sum_{m_1\cdots m_{l-1}m_l} P_t(m_1\cdots m_{l-1}m_l) 
\label{length_prob}
\ee
is the probability that the copolymer has the length $l$ at the time $t$, and $\mu(m_1\cdots m_{l-1}m_{l})$ is the stationary probability to find the sequence $\omega=m_1\cdots m_{l-1}m_{l}$ given that the length is equal to $l$.

By summing over the monomeric units composing the beginning of the copolymer, we define the probabilities:
\begin{subequations}
\bea
\mu(m_l) &\equiv& \sum_{m_1\cdots m_{l-1}} \mu(m_1\cdots m_{l-1}m_l) \, , \\
\mu(m_{l-1}m_l) &\equiv& \sum_{m_1\cdots m_{l-2}} \mu(m_1\cdots m_{l-1}m_l) \, ,\\
&\vdots& \nonumber
\eea
\end{subequations}
They are interpreted as follows:  $\mu(m_l)$ is the probability that $m_l$ is the last monomeric unit at the tip of the copolymer; $\mu(m_{l-1}m_l)$ is the probability that its last monomeric unit is $m_l$, and its penultimate unit is $m_{l-1}$; etc.  These probabilities are normalized and related to each other according to
\begin{subequations}
\label{norm}
\bea
\sum_{m_l}\mu(m_l) &=& 1 \, ,\label{norm1}\\
\sum_{m_{l-1}}\mu(m_{l-1}m_l) &=& \mu(m_l) \label{mu-mu} \, ,\label{norm2}\\
&\vdots&  \nonumber
\eea
\end{subequations}

Inserting the assumption~(\ref{Hyp}) into the kinetic equations~(\ref{kinetic_eqs}) and summing over all the monomeric units $m_1\cdots m_{l-1}m_l$, we find that
\be
\frac{d}{dt}\, p_t(l) = a \, p_t(l-1) +b \, p_t(l+1) - (a+b) \, p_t(l)
\label{eq-pl}
\ee
with the coefficients
\bea
a &=& \sum_{m_{l-1}m_l}  w_{+m_l\vert m_{l-1}} \, \mu(m_{l-1}) \, ,\\
b &=& \sum_{m_{l-1}m_l}  w_{-m_l\vert m_{l-1}} \, \mu(m_{l-1}m_l) \, .
\eea
Note that these coefficients depend on the distribution of monomers at the tip of the copolymer,  which will be deduced afterwards. 

Since Eq.~(\ref{eq-pl}) is linear, its general solution can be written as a linear superposition of solutions of the form $p_t(l)= \exp(s_q t + iql)$ with an arbitrary parameter $-\pi < q \leq +\pi$.  The exponential rate $s_q$ is here expected to take the form:
\be
s_q = - i \, q \, v - {\cal D}\, q^2 + O(q^3) \, ,
\label{dispersion1}
\ee
where $v$ is the mean growth velocity of the copolymer counted in monomers per second and $\cal D$ is the associated diffusivity in units of (monomers)$^2$ per second.  After a long enough time, the molar-mass dispersity (defined\cite{GHJJKS09} as the ratio of the mass-average molar mass\cite{F53} $\bar{M}_{\rm w}$ to the number-average molar mass\cite{F53} $\bar{M}_{\rm n}$) should thus tend to the unit value for such a living copolymerization process because $\bar{M}_{\rm w}/\bar{M}_{\rm n} \simeq \langle l^2\rangle_t/\langle l\rangle_t^2 \to 1$ in the limit $t\to\infty$.  With Eq.~(\ref{eq-pl}), we get
\be
s_q = (a+b)\, (\cos q-1)- i \,(a-b) \,\sin q \, ,
\label{dispersion2}
\ee
so that the mean velocity is given by $v=a-b$ and the diffusivity by ${\cal D}=(a+b)/2$.  Consequently, the mean velocity can be expressed as
\be
v = \sum_{m_{l-1}m_l} \left[ w_{+m_l\vert m_{l-1}} \, \mu(m_{l-1}) - w_{-m_l\vert m_{l-1}} \, \mu(m_{l-1}m_l)\right]  .
\label{velo0}
\ee

In order to determine the stationary probabilities, we insert again the assumption~(\ref{Hyp}) into the kinetic equations~(\ref{kinetic_eqs}), but we now sum over all the monomeric units except the last unit, thereafter, over all except the last and penultimate units, etc... After the substitution $p_t(l)= \exp(s_q t + iql)$ and taking the limit $q\to 0$, we obtain the following equations:
\begin{subequations}
\bea
0 &=& \sum_{m_{l-1}}  w_{+m_l\vert m_{l-1}} \, \mu(m_{l-1}) 
+\sum_{m_{l+1}} w_{-m_{l+1}\vert m_{l}} \, \mu(m_{l}m_{l+1}) \nonumber\\
&-& \sum_{m_{l-1}} w_{-m_{l}\vert m_{l-1}} \, \mu(m_{l-1}m_l) 
- \sum_{m_{l+1}} w_{+m_{l+1}\vert m_{l}}\, \mu(m_{l}) \, ,\label{eq-mu1}\\
0 &=& w_{+m_l\vert m_{l-1}} \, \mu(m_{l-1}) 
+\sum_{m_{l+1}} w_{-m_{l+1}\vert m_{l}} \, \mu(m_{l-1}m_{l}m_{l+1}) \nonumber\\
&-& \left( w_{-m_{l}\vert m_{l-1}} + \sum_{m_{l+1}} w_{+m_{l+1}\vert m_{l}}\right) \mu(m_{l-1}m_l) \, ,\label{eq-mu2}\\
0 &=& w_{+m_l\vert m_{l-1}} \, \mu(m_{l-2}m_{l-1}) 
+\sum_{m_{l+1}} w_{-m_{l+1}\vert m_{l}} \, \mu(m_{l-2}m_{l-1}m_{l}m_{l+1}) \nonumber\\
&-& \left( w_{-m_{l}\vert m_{l-1}} + \sum_{m_{l+1}} w_{+m_{l+1}\vert m_{l}}\right) \mu(m_{l-2}m_{l-1}m_l) \, ,\label{eq-mu3}\\
&&\qquad\qquad\qquad\qquad\qquad\vdots \nonumber
\eea
\end{subequations}
These equations are related to each other by Eq.~(\ref{mu-mu}) and the next ones.
Indeed, summing Eq.~(\ref{eq-mu3}) over $m_{l-2}$, we get Eq.~(\ref{eq-mu2});
summing Eq.~(\ref{eq-mu2}) over $m_{l-1}$, we get Eq.~(\ref{eq-mu1}); 
and summing Eq.~(\ref{eq-mu1}) over $m_{l}$, we get an equation that is trivially satisfied.
Therefore, all these equations are consistent with each other and also with the normalization conditions (\ref{norm}).  A crucial observation is that, in this hierarchy, all the equations except the first one Eq.~(\ref{eq-mu1}) have the same structure with the same coefficients for the same last monomeric units.  
This structure suggests to look for a solution of the form
\begin{subequations}
\bea
\mu(m_{l-1}m_{l}) &=& \mu(m_{l-1}\vert m_{l}) \, \mu(m_{l}) \, ,\label{mu-mu2} \\
\mu(m_{l-2}m_{l-1}m_{l}) &=& \mu(m_{l-2}\vert m_{l-1}) \, \mu(m_{l-1}\vert m_{l}) \, \mu(m_{l}) \, ,\label{mu-mu3}\\
&\vdots& \nonumber
\eea
\end{subequations}
in terms of the {\it conditional probabilities} $\mu(m_{l-1}\vert m_{l})$ satisfying
\be
\sum_{m_{l-1}} \mu(m_{l-1}\vert m_{l})=1 \, ,
\label{norm_c}
\ee
and the {\it tip probabilities} $\mu(m_l)$ for the last monomeric unit $m_l$ composing the chain of length $l$.

Now, we need to determine these conditional and tip probabilities.  Replacing Eqs.~(\ref{mu-mu2}) and (\ref{mu-mu3}) into Eqs.~(\ref{eq-mu1}) and (\ref{eq-mu2}), we obtain the following coupled nonlinear equations: 
\bea
\mu(m_{l-1}\vert m_l) &=& \frac{w_{+m_l\vert m_{l-1}} \, \mu(m_{l-1})}{
\left( w_{-m_{l}\vert m_{l-1}} + \sum_{m_{l+1}} w_{+m_{l+1}\vert m_{l}}\right)\mu(m_{l}) -\sum_{m_{l+1}} w_{-m_{l+1}\vert m_{l}} \, \mu(m_l\vert m_{l+1}) \, \mu(m_{l+1})} \, ,\label{cond_proba}\\
\mu(m_{l}) &=& \frac{\sum_{m_{l-1}}  w_{+m_l\vert m_{l-1}} \, \mu(m_{l-1}) 
+\sum_{m_{l+1}} w_{-m_{l+1}\vert m_{l}} \, \mu(m_{l}\vert m_{l+1}) \, \mu(m_{l+1})}{\sum_{m_{l+1}} w_{+m_{l+1}\vert m_{l}}+\sum_{m_{l-1}} w_{-m_{l}\vert m_{l-1}} \, \mu(m_{l-1}\vert m_l) } \, .\label{tip_proba}
\eea
The mean velocity can in turn be expressed as
\be
v = \sum_{m_{l-1}m_l} \left[ w_{+m_l\vert m_{l-1}} \, \mu(m_{l-1}) - w_{-m_l\vert m_{l-1}} \, \mu(m_{l-1}\vert m_l)\, \mu(m_l)\right]  .
\label{velo1}
\ee
Since there are $M^2$ equations~(\ref{cond_proba}) for the $M^2$ conditional probabilities and $M$ equations~(\ref{tip_proba}) for the $M$ tip probabilities $\mu(m_{l})$, we have a complete set of equations for these $M(M+1)$ probabilities.  

\subsection{The first-order Markov chain}

The previous results show that the stationary distribution of sequences is described by a first-order Markov chain according to
\be
\mu(m_1m_2\cdots m_{l-1}m_l) = \mu(m_1\vert m_2) \cdots \mu(m_{l-1}\vert m_l) \, \mu(m_l) \, .
\label{MarkovChain}
\ee
We notice that this Markov chain runs `backwards', i.e., it describes the copolymer structure starting from the tip of the growing copolymer. 

The conditional and tip probabilities are given by the coupled Eqs.~(\ref{cond_proba}) and (\ref{tip_proba}).  However, these equations are complicated and difficult to use.  
To reduce this complexity,
we introduce the partial velocities
\be
v_m \equiv \sum_{n=1}^M w_{+n\vert m} - \frac{1}{\mu(m)} \, \sum_{n=1}^M w_{-n\vert m} \, \mu(m \vert n) \, \mu(n)
\label{partial_velo}
\ee
for $m=1,2,...,M$.  Inserting these velocities into Eqs.~(\ref{cond_proba}), the conditional probabilities can be written as
\be
\mu(m\vert n) = \frac{w_{+ n\vert m} \, \mu(m)}{(w_{-n\vert m} + v_n) \, \mu(n)}
\label{cond_proba_2}
\ee
for $m,n=1,2,...,M$.  Replacing these relations into Eqs.~(\ref{partial_velo}), we obtain the following self-consistent equations for the partial velocities:
\be
v_m = \sum_{n=1}^M \frac{w_{+ n\vert m} \, v_n}{w_{-n\vert m} + v_n}
\label{central_eqs}
\ee
for $m=1,2,...,M$.  This system of $M$ equations can be solved numerically
starting from positive values for the partial velocities.  In this way, the partial velocities can be directly obtained from the knowledge of the reaction rates $w_{\pm m\vert n}$.

Now, multiplying Eqs.~(\ref{cond_proba_2}) by $\mu(n)$, summing over $m$ and using the normalization conditions (\ref{norm_c}), we find the following linear system of equations for the tip probabilities:
\be
\sum_{m=1}^M \frac{w_{+ n\vert m}}{w_{-n\vert m} + v_n} \, \mu(m) = \mu(n)
\label{tip_proba_2}
\ee
with $n=1,2,...,M$.
Finally, the mean velocity (\ref{velo1}) can be expressed as
\be
v=\sum_{m=1}^M v_m \, \mu(m) \, .
\label{velo2}
\ee

In summary, once the partial velocities are found by solving Eqs.~(\ref{central_eqs}), we successively obtain the tip probabilities with Eqs.~(\ref{tip_proba_2}), the conditional probabilities with Eqs.~(\ref{cond_proba_2}), and the mean growth velocity with Eq.~(\ref{velo2}).  We notice that Eqs.~(\ref{central_eqs}) may admit several solutions, but the conditional probabilities~(\ref{cond_proba_2}), the tip probabilities~(\ref{tip_proba_2}), and the mean velocity~(\ref{velo2}) of the solution we are looking for must be positive.  In Section \ref{Thermo}, we derive the conditions for the attachment and detachment rates under which copolymerization occurs.\\

The properties of the Markov chain (\ref{MarkovChain}) can be deduced from the conditional and tip probabilities.  In particular, the probability to find any type of monomeric unit at some distance behind the tip can now be determined.  The probability to find $m_{l-1}$ as the penultimate monomeric unit is given by 
\be
\mu^{(-1)}(m_{l-1})=\sum_{m_l}\mu(m_{l-1}\vert m_l) \, \mu(m_l)\, ,
\ee
the probability to find the unit $m_{l-2}$ at the previous position by
\be
\mu^{(-2)}(m_{l-2}) = \sum_{m_{l-1}m_l}\mu(m_{l-2}\vert m_{l-1})\, \mu(m_{l-1}\vert m_l) \, \mu(m_l)\, ,
\ee
and so on towards the beginning of the copolymer sequence. In general, the probability to find the unit $m_{l-k}$ at the $k^{\rm th}$ position behind the tip unit $m_l$ can be written as
\be
\mu^{(-k)}(m_{l-k}) = \sum_{m_l}\left({\boldsymbol{\mathsf M}}^k\right)_{m_{l-k}m_l} \, \mu(m_l)\, ,
\label{mu_k}
\ee
in terms of the matrix of conditional probabilities: 
\be
({\boldsymbol{\mathsf M}})_{mn}=\mu(m\vert n)\, .
\label{matrix-M}
\ee

Suppose that this matrix can be decomposed into its eigenvalues and associated eigenvectors satisfying
\be
\left\{
\begin{array}{l}
\sum_{m=1}^M \xi_\alpha(m) \, \mu(m\vert n) = \Lambda_\alpha \, \xi_\alpha(n)\, , \\
\sum_{n=1}^M\mu(m\vert n) \, \eta_\alpha(n) = \Lambda_\alpha \, \eta_\alpha(m) \, ,
\end{array}
\right.
\label{eigenvalue_problem}
\ee
with $\alpha=1,2,...,M$.  The left and right eigenvectors should satisfy the biorthonormality condition, $\sum_{m=1}^M\xi_\alpha(m)\, \eta_\beta(m)=\delta_{\alpha\beta}$.  Because of the normalization conditions (\ref{norm_c}), the leading eigenvalue is $\Lambda_1=1$, while the corresponding eigenvectors are $\xi_1(m)=1$ and $\eta_1(m)$, which defines a probability distribution.  The probability~(\ref{mu_k}) to find $m_{l-k}$ as the $k^{\rm th}$ monomeric unit behind the tip unit can then be decomposed as:
\be
\mu^{(-k)}(m_{l-k}) = \eta_1(m_{l-k})+\sum_{\alpha\neq 1} \eta_\alpha(m_{l-k}) \, \left(\Lambda_\alpha\right)^k \sum_{m_l} \xi_\alpha(m_l)\, \mu(m_l)\, .
\label{mu_k-2}
\ee
Since the eigenvalues with $\alpha\neq 1$ are smaller than unity in absolute value $\vert\Lambda_\alpha\vert < 1$, the probability (\ref{mu_k-2}) converges exponentially towards $\eta_1(m_{l-k})$ as $k$ increases.  Hence, the right eigenvector $\eta_1(m)=\bar\mu(m)$ associated with the leading eigenvalue $\Lambda_1=1$ gives the probability to find the monomeric unit $m$ in the bulk of the sequence:
\be
\bar{\mu}(m)=\sum_{n=1}^M\mu(m\vert n) \, \bar{\mu}(n) \, .
\label{bulk_proba}
\ee
The solution of the system~(\ref{bulk_proba}) can be expressed in terms of the tip probabilities and the partial velocities as
\be
\bar{\mu}(m) = \frac{\mu(m) v_m}{v} \, .
\label{bulk_proba_2}
\ee
Thus, the bulk probabilities $\bar\mu(m)$ generally differ from the tip probabilities $\mu(m)$. They only coincide when all partial velocities take the same value, $v_m = v$.

The eigenvalues $\Lambda_\alpha$ also determine the decay of correlation functions characterizing the statistics of the monomeric units in the sequence.  Such a correlation function can be defined as
\be
C(j) =  \left\langle \left( m_i -\langle m\rangle \right) \left( m_{i+j} -\langle m\rangle \right)\right\rangle
\label{correl}
\ee
where the statistical average $\langle\cdot\rangle$ is carried out over an ensemble of sequences generated by copolymerization.  The average of the index of the monomeric unit takes the value $\langle m\rangle=\sum_{m=1}^M m\, \bar\mu(m)$ and the correlation function can be written as
\be
C(j) =  \sum_{m,n=1}^M (m-\langle m\rangle) \left({\boldsymbol{\mathsf M}}^j\right)_{mn} \, \bar\mu(n) \, (n-\langle m\rangle) \, ,
\label{correl-matrix}
\ee
with the matrix (\ref{matrix-M}) of the conditional probabilities. Since this matrix can be decomposed in terms of its eigenvalues and eigenvectors~(\ref{eigenvalue_problem}), the correlation function decays to zero as
\be
C(j) = \sum_{\alpha\neq 1} c_{m\alpha} \Lambda_{\alpha}^j
\label{correl-eigen}
\ee
with some coefficients $c_{m\alpha}$.  This result also holds for other correlation functions defined by replacing $m_i$ and $m_{i+j}$ by any functions $A(m_i)$ and $B(m_{i+j})$ in Eq.~(\ref{correl}): $C_{AB}(j) =  \left\langle [ A(m_i) -\langle A\rangle ] [B(m_{i+j}) -\langle B\rangle]\right\rangle$.

If the matrix~(\ref{matrix-M}) could not be decomposed into eigenvalues and eigenvectors, a Jordan form decomposition would be required.\cite{M00}  In the presence of Jordan blocks, correlation functions decay more slowly than exponentially: $C(j) \simeq j^{d_\alpha-1} \Lambda_{\alpha}^j $, where $d_{\alpha}$ is the multiplicity of the eigenvalue $\Lambda_{\alpha}$.

\subsection{Entropy production}

As shown in Refs.~\onlinecite{AG08} and~\onlinecite{AG09}, the thermodynamic entropy production is given by
\be
\frac{1}{k_{\rm B}} \frac{d_{\rm i}S}{dt} = v \, A = v \, (\epsilon+D) \geq 0
\label{entrprod}
\ee
where $k_{\rm B}$ is Boltzmann's constant, $v$ the mean velocity~(\ref{velo2}), and $A$ the associated affinity.  This latter is the sum of the free-energy driving force $\epsilon$ and the Shannon disorder per monomer $D$.  According to the second law of thermodynamics, the entropy production is always non negative.  The free-energy driving force is determined by the free enthalpy per monomer $g$ in the copolymer sequence and the temperature $T$ as
\be
\epsilon = -\frac{g}{k_{\rm B}T} = \sum_{m,n=1}^M \bar\mu(n) \, \mu(m\vert n) \, \ln \frac{w_{+n\vert m}}{w_{-n\vert m}} \, .
\label{eps}
\ee
For a first-order Markov chain, the Shannon disorder per monomer is given by
\be
D = - \sum_{m,n=1}^M \bar\mu(n) \, \mu(m\vert n) \, \ln \mu(m\vert n) \ge 0 \, ,
\label{disorder}
\ee
which is always non-negative.\cite{CT06}  This quantity is bounded from above by the number of monomeric types: $D\leq \ln M$.

There exist two regimes of growth for the copolymer: (1) the regime driven by free energy, if $\epsilon >0$; (2) the regime driven by the entropic effect of the disorder in the sequence of the growing copolymer, if $-D<\epsilon<0$.  In both regimes, the affinity $A=\epsilon+D$ is positive so that the velocity $v$ is positive in accordance with the second law (\ref{entrprod}).\cite{AG08,AG09,J08}

\subsection{The fully irreversible regime}

The fully irreversible regime of copolymerization is reached if the detachment rates are vanishing:
\be
w_{-m\vert n}=0 \, .
\ee
In this regime, the free-energy driving force~(\ref{eps}) and the thermodynamic entropy production~(\ref{entrprod}) become infinite.  

Considering the case of two monomers $M=2$, Eqs.~(\ref{partial_velo})-(\ref{velo2}) allow us to obtain the growth velocity and the composition of the growing copolymer in terms of the reactivity ratios
\be
r_1 \equiv \frac{k_{+1\vert 1}}{k_{+2\vert 1}} \, , \qquad r_2 \equiv \frac{k_{+2\vert 2}}{k_{+1\vert 2}} \, ,
\ee
and the mole fractions of the monomers in the feed:
\be
f_1 \equiv \frac{[1]}{[1]+[2]} \, , \qquad f_2 \equiv \frac{[2]}{[1]+[2]} \, .
\ee
The partial velocities~(\ref{partial_velo}) are the sums of the attachment rates: $v_m=\sum_{n=1}^M w_{+n\vert m}$.  According to Eqs.~(\ref{tip_proba_2}), the tip probabilities are given by
\begin{subequations}
\bea
\mu(1) &=& \frac{w_{+1\vert 2}}{w_{+1\vert 2}+w_{+2\vert 1}} \, , \\
\mu(2) &=& \frac{w_{+2\vert 1}}{w_{+1\vert 2}+w_{+2\vert 1}} \, , 
\eea
\end{subequations}
their ratio by
\be
\frac{\mu(1)}{\mu(2)} = \frac{w_{+1\vert 2}}{w_{+2\vert 1}} =\frac{k_{+1\vert 2} \, f_1}{k_{+2\vert 1} \, f_2} \, ,
\ee
and the conditional probabilities~(\ref{cond_proba_2}) by
\begin{subequations}
\bea
\mu(1\vert 1) &=& \frac{w_{+1\vert 1}}{w_{+1\vert 1}+w_{+2\vert 1}} = \frac{r_1\, f_1}{r_1\, f_1+f_2} \, , \\
\mu(2\vert 1) &=& \frac{w_{+2\vert 1}}{w_{+1\vert 1}+w_{+2\vert 1}} = \frac{f_2}{r_1\, f_1+f_2} \, , \\
\mu(1\vert 2) &=& \frac{w_{+1\vert 2}}{w_{+1\vert 2}+w_{+2\vert 2}} = \frac{f_1}{f_1+r_2\, f_2} \, , \\
\mu(2\vert 2) &=& \frac{w_{+2\vert 2}}{w_{+1\vert 2}+w_{+2\vert 2}} = \frac{r_2\, f_2}{f_1+r_2\, f_2} \, ,
\eea
\end{subequations}
in agreement with Ref.~\onlinecite{AG44}.  Consequently, the mean growth velocity~(\ref{velo1}) can be expressed as
\be
v = (r_1\, f_1^2 + 2\, f_1 \, f_2 + r_2\, f_2^2) \, \frac{k_{+1\vert 2} \, k_{+2\vert 1}}{k_{+1\vert 2} \, f_1 + k_{+2\vert 1} \, f_2} \, .
\ee

Finally, the ratio of the mole fractions of monomeric units in the copolymer, $F_1\equiv\bar\mu(1)$ and $F_2\equiv\bar\mu(2)$, is obtained as
\be
\frac{F_1}{F_2} = \frac{\bar\mu(1)}{\bar\mu(2)} = \frac{f_1(r_1\, f_1+f_2)}{f_2(f_1+r_2\, f_2)}\, ,
\ee
which is the Mayo-Lewis formula.\cite{ML44}  The classic results of Refs.~\onlinecite{ML44} and~\onlinecite{AG44} are thus recovered in the fully irreversible regime.

\section{Thermodynamic equilibrium}
\label{Thermo}

\subsection{Detailed balance}

The state of thermodynamic equilibrium can be identified by the conditions of detailed balancing between the reactions of attachment and detachment of monomers in the kinetic equations (\ref{kinetic_eqs}):
\be
w_{+m_l\vert m_{l-1}} \, P_{\rm eq}(m_1\cdots m_{l-1})= w_{-m_{l}\vert m_{l-1}} \,  P_{\rm eq}(m_1\cdots m_{l-1}m_{l}) \, .
\label{detailed_balance}
\ee
Summing over the monomeric units $m_1\cdots m_{l-2}$, these conditions read $w_{+m_l\vert m_{l-1}} \mu_{\rm eq}(m_{l-1})= w_{-m_{l}\vert m_{l-1}} \mu_{\rm eq}(m_{l-1}m_{l})$.  In terms of the conditional and tip probabilities $\mu_{\rm eq}(m_{l-1}\vert m_{l})$ and $\mu_{\rm eq}(m_{l})$, we find that
\be
\mu_{\rm eq}(m\vert n) = \frac{w_{+n\vert m}\, \mu_{\rm eq}(m)}{w_{-n\vert m}\, \mu_{\rm eq}(n)}
\label{equil_cond}
\ee
for $m,n=1,2,...,M$, at equilibrium.  By replacing into Eqs.~(\ref{partial_velo}), we can verify that the partial velocities vanish at equilibrium.  Hence, the mean velocity (\ref{velo2}) also vanishes at equilibrium.
By the normalization conditions (\ref{norm_c}) for the conditional probabilities $\mu_{\rm eq}(m\vert n)$, the tip probabilities should satisfy
\be
\sum_{m=1}^M z_{n\vert m} \, \mu_{\rm eq}(m) = \mu_{\rm eq}(n)
\ee
with
\be 
z_{n\vert m} \equiv \frac{w_{+n\vert m}}{w_{-n\vert m}} \, ,
\label{z}
\ee
whereupon the necessary and sufficient condition for the system to be at equilibrium is that the spectral radius of
the $M\times M$ matrix
\be
{\boldsymbol{\mathsf Z}} = (z_{n\vert m})
\label{mz}
\ee
is equal to unity: 
\be
\rho({\boldsymbol{\mathsf Z}}) = 1 \, .
\label{NSC_eq}
\ee
The spectral radius of a matrix is defined as the maximum of the absolute values of its eigenvalues $z_\alpha$:
$\rho({\boldsymbol{\mathsf Z}})\equiv{\rm max}_{\alpha}\{\vert z_\alpha\vert\}$. The condition~(\ref{NSC_eq}) implies that the leading eigenvalue of the matrix~(\ref{mz}) is equal to unity so that
\be
\det({\boldsymbol{\mathsf Z}}-{\boldsymbol{\mathsf 1}})=0 
\label{Z-eq}
\ee
at equilibrium.

We notice that the copolymer is growing if $\rho({\boldsymbol{\mathsf Z}})>1$ and depolymerizing if $\rho({\boldsymbol{\mathsf Z}})<1$.

\subsection{Entropy production}

At equilibrium, not only the mean velocity but also the affinity $A=\epsilon+D$ vanishes. Indeed, using the equilibrium conditions (\ref{equil_cond}), the free-energy driving force is related to the Shannon disorder according to
\be
\epsilon_{\rm eq} = - D_{\rm eq}
\label{eps-D-equil}
\ee
and the entropy production~(\ref{entrprod}) vanishes at equilibrium.

\section{Depolymerization}
\label{Depolym}

\subsection{The initial copolymer}

The concentrations of monomers in the solution may be such that a copolymer will depolymerize because the detachment  of monomeric units overwhelms their attachment.  Precisely, depolymerization occurs when the concentrations of monomers are such that $\rho({\boldsymbol{\mathsf Z}}) < 1$.
This process is also ruled by the kinetic equations~(\ref{kinetic_eqs}), but with a different initial condition.  In this case, at the initial time $t=0$, the solution contains a copolymer that has been synthesized under conditions different from those that prevail in the solution.  Therefore, the initial sequence of the copolymer is arbitrary: $\omega_0=m_1m_2\cdots m_{l-1}m_l$.  The initial probability distribution is thus given by $P_0(\omega)=1$ if $\omega=\omega_0$ and $P_0(\omega)=0$ otherwise.  

The initial copolymer $\omega_0$ is characterized by the occurrence frequencies of its monomeric units $m_j$, dyads $m_jm_{j+1}$, triads $m_jm_{j+1}m_{j+2}$,...
\be
\bar\mu_1(m), \, \bar\mu_2(mm'),\, \bar\mu_3(mm'm''),\, \dots
\ee
These frequencies are normalized and related to each other according to
\begin{subequations}
\bea
\sum_{m=1}^M \bar\mu_1(m) &=& 1 \, ,\\
\sum_{m'=1}^M \bar\mu_2(mm') &=& \bar\mu_1(m) \, ,\\
&\vdots&\nonumber
\eea
\end{subequations}
All the statistical properties of the initial copolymer are determined by these frequencies.  In particular, the possible information content of the sequence $\omega_0$ is given by the Shannon information per monomer as
\be
\bar I_\infty = \lim_{l\to\infty} - \frac{1}{l} \sum_{m_1\cdots m_l} \bar\mu_l(m_1\cdots m_l) \, \ln  \bar\mu_l(m_1\cdots m_l) \, .
\ee
This information is erased during depolymerization.\cite{AG13}

\subsection{The mean depolymerization velocity}

During the depolymerization of $\omega_0$, monomeric units are progressively removed from the chain so that the mean velocity is here negative $v<0$.  If the attachment rates are not all vanishing, attachment events create transient growth phases even though the polymer's length decreases on average.\cite{AG13}  This process determines the mean depolymerization velocity, as the following reasoning shows. 

In the regime of depolymerization, the copolymer will eventually return with probability one to the sequence it had before a transient growth phase.  Back to the sequence $m_1\cdots m_l$, the copolymer will remove the monomer $m_l$ with some probability, otherwise it will start growing again.  We consider the mean time $\langle T_{m_l\vert m_{l-1}}\rangle$ before the monomeric unit $m_l$ is detached from the penultimate unit $m_{l-1}$.  This removal may be preceded by transient growth phases during which $r$ monomeric units are added.  Accordingly, the removal of $m_l$ is a first-passage problem.  If $r=0$, the mean time to remove $m_l$ is equal to $(w_{-m_l\vert m_{l-1}})^{-1}$.  If $r=1$, the transitions occurring before removal are $m_l\to m_lm_{l+1}\to m_l$, which takes the mean time $(w_{-m_l\vert m_{l-1}})^{-1}\sum_{m_{l+1}}w_{+m_{l+1}\vert m_{l}}/w_{-m_{l+1}\vert m_{l}}$.  Considering all the possible values of $r$, the mean time to remove the monomer $m_l$ is given by
\be
\langle T_{m_l\vert m_{l-1}}\rangle = \frac{1}{w_{-m_l\vert m_{l-1}}} \sum_{r=0}^{\infty} \sum_{m_{l+1}\cdots m_{l+r}} \prod_{i=l}^{l+r-1} z_{m_{i+1}\vert m_i} \, ,
\ee
in terms of the ratios $z_{n\vert m}\equiv w_{+n\vert m}/w_{-n\vert m}$ defined in Eq.~(\ref{z}). Using the matrix notation~(\ref{mz}), the mean time reads
\be
\langle T_{m_l\vert m_{l-1}}\rangle = \frac{1}{w_{-m_l\vert m_{l-1}}} \sum_{r=0}^{\infty} \sum_{n=1}^M \left({\boldsymbol{\mathsf Z}}^r\right)_{nm_l}= \frac{1}{w_{-m_l\vert m_{l-1}}} \sum_{n=1}^M \left(\frac{\boldsymbol{\mathsf 1}}{{\boldsymbol{\mathsf 1}}-{\boldsymbol{\mathsf Z}}}\right)_{nm_l} \, .
\label{mean_time}
\ee
Note that the matrix ${\boldsymbol{\mathsf 1}}/({\boldsymbol{\mathsf 1}}-{\boldsymbol{\mathsf Z}})$ is well defined as $\rho({\boldsymbol{\mathsf Z}}) < 1$ in the depolymerization regime.

During the depolymerization, this process will be repeated over the successive monomers of the original chain.  Therefore, the mean depolymerization velocity is obtained by averaging the mean times (\ref{mean_time}) over the dyad frequencies as
\be
-v = \left[\sum_{m,m'=1}^M \bar\mu_2(m m') \, \langle T_{m'\vert m}\rangle\right]^{-1} \, .
\ee
Inserting the expression (\ref{mean_time}) for the mean times, the mean velocity becomes
\be
v = - \left[\sum_{m,m'=1}^M  \frac{\bar\mu_2(mm')}{w_{-m'\vert m}} \sum_{n=1}^M \left(\frac{\boldsymbol{\mathsf 1}}{{\boldsymbol{\mathsf 1}}-{\boldsymbol{\mathsf Z}}}\right)_{nm'}\right]^{-1} \, .
\label{mean_velo}
\ee
We notice that this velocity depends on the dyad frequency in agreement with the fact that the process only involves the previous monomeric unit.  At equilibrium, the depolymerization velocity~(\ref{mean_velo}) vanishes because the determinant~(\ref{Z-eq}) vanishes.

\subsection{Entropy production}
\label{VC}

During depolymerization, the copolymer sequence does not differ from the initial sequence except during the transient growth phases.  However, this addition remains of finite length with respect to the length of the initial sequence, which is supposed to be very long for depolymerization to proceed in a steady regime. Therefore, the entropy production has no contribution due to the proliferation of possible sequences as it was the case during copolymerization and the Shannon disorder per monomer is thus equal to zero.  The only contribution to the entropy production is due to the free enthalpy of depolymerization\cite{AG13}
\be
\frac{1}{k_{\rm B}} \frac{d_{\rm i}S}{dt} = v \, \bar\epsilon \geq 0 \, .
\label{entrprod-depolym}
\ee
The free-energy driving force is here given by
\be
\bar\epsilon = -\frac{\bar g}{k_{\rm B}T} = \sum_{m,n=1}^M \bar\mu_2(mn) \, \ln \frac{w_{+n\vert m}}{w_{-n\vert m}} \, .
\label{eps-depolym}
\ee
Since the free enthalpy per monomer is positive during depolymerization $\bar g>0$, the mean velocity must be negative, as expected from the second law.  In general, this free enthalpy is bounded from below by the Shannon information per monomer\cite{AG13}
\be
\bar g \ge k_{\rm B}T \, \bar I_\infty \, .
\label{Binfty}
\ee
The erasure of information  contained in the initial copolymer should thus dissipate energy in agreement with Landauer's principle.\cite{AG13}  For the class of processes that only depend on the previous monomeric unit, the lower bound under the constraint of zero velocity is given by
\be
\bar g \ge k_{\rm B}T \, \bar I_2 \, ,
\label{B2}
\ee
in terms of the Shannon information in the dyads
\be
\bar I_2 = - \sum_{m,n=1}^M \bar\mu_2(mn) \, \ln  \bar\mu_2(m\vert n) \, ,
\label{I2}
\ee
with $\bar\mu_2(m\vert n)\equiv\bar\mu_2(mn)/\bar\mu_1(n)$.
The inequality (\ref{B2}) is obtained by considering the Kullback-Leibler divergence between the probability distribution of the first-order Markov chain generated by the conditional probabilities $\bar\mu_2(m\vert n)$ and the one generated by Eq.~(\ref{equil_cond}) at equilibrium where the constraint of zero velocity is satisfied:
\be
D_{\rm KL}\left(\bar\mu_2\Vert\mu_{\rm eq}\right) = \sum_{m,n=1}^M \bar\mu_2(mn) \, \ln\frac{\bar\mu_2(m\vert n)}{\mu_{\rm eq}(m\vert n)} \geq 0 \, ,
\ee
which is known to be always non negative.\cite{CT06}  Replacing the equilibrium distribution by its expression (\ref{equil_cond}), the Kulbback-Leibler divergence turns out to be equal to the difference between the free enthalpy per monomer in units of the thermal energy and the Shannon information~(\ref{I2}):
\be
D_{\rm KL}\left(\bar\mu_2\Vert\mu_{\rm eq}\right) = \frac{\bar g}{k_{\rm B}T} - \bar I_2 \geq 0 \, ,
\ee
hence the inequality (\ref{B2}).  The equality $\bar g=k_{\rm B}T \, \bar I_2$ would hold if the initial copolymer had precisely the equilibrium composition: $\bar\mu_2(m\vert n)=\mu_{\rm eq}(m\vert n)$.  However, the initial copolymer is usually grown under quite different conditions so that the equality is typically not met for the depolymerization of a given initial copolymer by a given process.

In general, the bound (\ref{Binfty}) is always satisfied because $\bar I_2\geq \bar I_\infty$.

\subsection{The fully irreversible regime}

The fully irreversible regime of depolymerization is reached when the attachment rates vanish:
\be
w_{+m\vert n}=0 \, .
\ee
In this case, the free-energy driving force~(\ref{eps-depolym}) and the thermodynamic entropy production~(\ref{entrprod-depolym}) become infinite.   

The matrix~(\ref{mz}) vanishes, ${\boldsymbol{\mathsf Z}}=0$, and the maximum depolymerization velocity is thus given by
\be
v = -\left[\sum_{m,m'=1}^M  \frac{\bar\mu_2(mm')}{w_{-m'\vert m}}\right]^{-1} \, .
\label{max_depolym_velo}
\ee

\section{Examples}
\label{Examples}

In this section, we consider several examples of copolymerization or depolymerization.
The stochastic processes are simulated with Gillespie's kinetic Monte-Carlo algorithm\cite{G76,G77} and the
results are compared with the theory.  In every figure, the dots depict data from simulations
and the solid lines the corresponding theoretical predictions.  The concentrations are given in moles per liter and the rates in (second)$^{-1}$.

\subsection{Copolymerization with $M=2$ monomers}

We start by investigating the properties of copolymer formation with $M=2$ monomers, $m\in\{1,2\}$, with attachment and detachment rate constants given by
\bea
&& k_{+1\vert 1} = 2 \, , \ k_{+1\vert 2} = k_{+2\vert 1} =k_{+2\vert 2} = 1 \, ,\nonumber\\
&& k_{-1\vert 1} = k_{-1\vert 2} = k_{-2\vert 1} =k_{-2\vert 2} = 0.01  \, , \nonumber\\
&& [2] = 0.01\, . \label{model1}
\eea
The concentration $[2]$ of monomer $m=2$ is kept fixed and the concentration $[1]$ of the other monomer $m=1$ is used as a control parameter.  Under the conditions~(\ref{model1}), the state of thermodynamic equilibrium (\ref{Z-eq}) is reached at $[1]_{\rm eq}=0$.

\begin{figure}[h]
\centerline{\scalebox{0.47}{\includegraphics{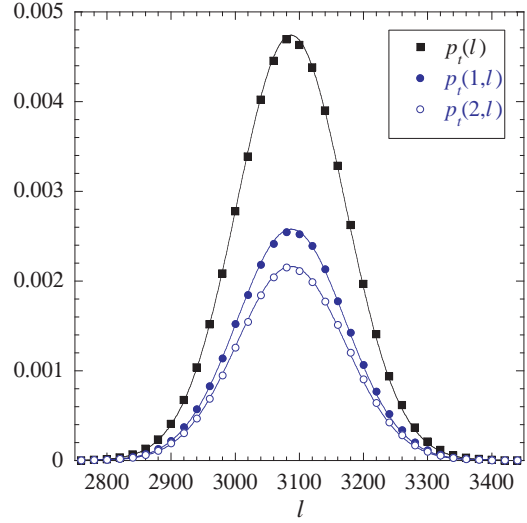}}}
\caption{Growth of a copolymer in the conditions (\ref{model1}) and $[1]=0.01$ until the time $t=200000$: the length probability $p_t(l)$ defined by Eq.~(\ref{length_prob}) (filled squares) and the probabilities $p_t(m;l)$ that the copolymer has the length $l$ and its last monomeric unit is $m=1$ or $m=2$ (filled and open circles) versus the copolymer length $l$.  The statistics is established with $N=10^7$ samples.  The solid lines show the theoretical predictions (\ref{ptl-diff}) and (\ref{ptml}) with the mean growth velocity $v=0.015437$, the diffusivity ${\cal D}=0.017718$, and the tip probabilities $\mu(1)=0.5437$ and $\mu(2)=0.4563$.  In the present conditions, the conditional probabilities are $\mu(1\vert 1)=0.7044$, $\mu(1\vert 2)=0.5437$, $\mu(2\vert 1)=0.2956$, and $\mu(2\vert 2)=0.4563$, the bulk probabilities $\bar\mu(1)=0.6478$ and $\bar\mu(2)=0.3522$, the Shannon disorder per monomer $D=0.6361$, and the free-energy driving force $\epsilon=0.3163$.  The average length is $\langle l\rangle_t=v\, t=3087.4$. }
\label{fig1}
\end{figure}

\begin{figure}[h]
\centerline{\scalebox{0.495}{\includegraphics{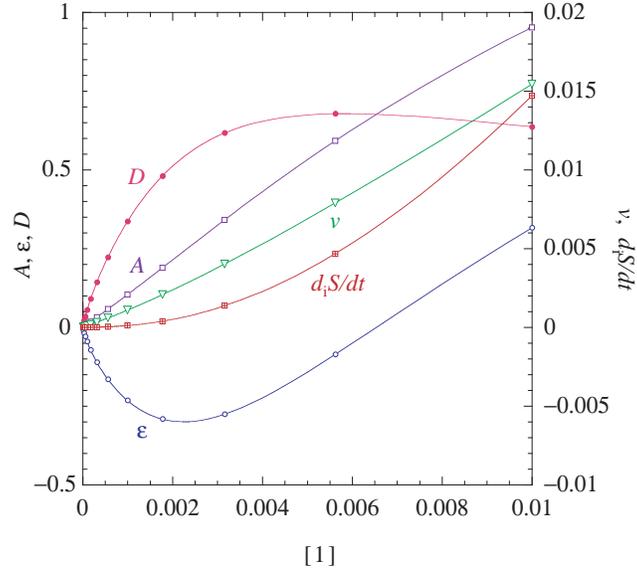}}}
\caption{Growth of a copolymer in the conditions (\ref{model1}):  The growth velocity $v$, the free-energy driving force $\epsilon$, the Shannon disorder per monomer $D$, the affinity $A=\epsilon+D$, and the thermodynamic entropy production $d_{\rm i}S/dt=A\, v$ with $k_{\rm B}=1$ versus the concentration $[1]$ of monomers $m=1$.  The dots are the results of simulations with Gillespie's algorithm, while the solid lines are the theoretical quantities obtained with Eqs.~(\ref{cond_proba_2})-(\ref{velo2}) and~(\ref{entrprod})-(\ref{disorder}).}
\label{fig2}
\end{figure}

\begin{figure}[h]
\centerline{\scalebox{0.5}{\includegraphics{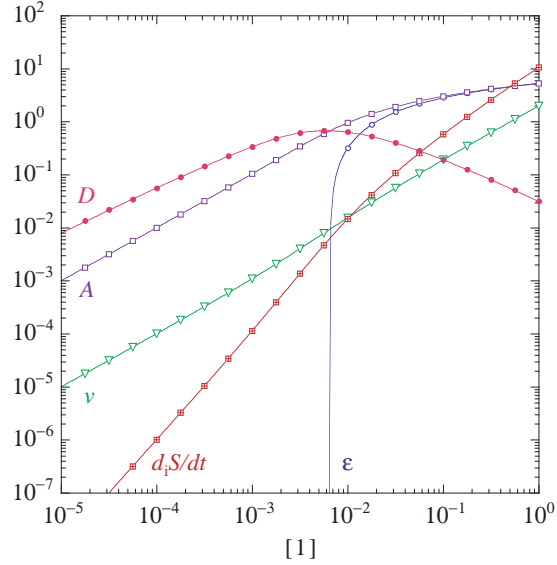}}}
\caption{Growth of a copolymer in the conditions (\ref{model1}): same quantities as in Fig.~\ref{fig2} plotted in logarithmic scales.}
\label{fig3}
\end{figure}

\begin{figure}[h]
\centerline{\scalebox{0.5}{\includegraphics{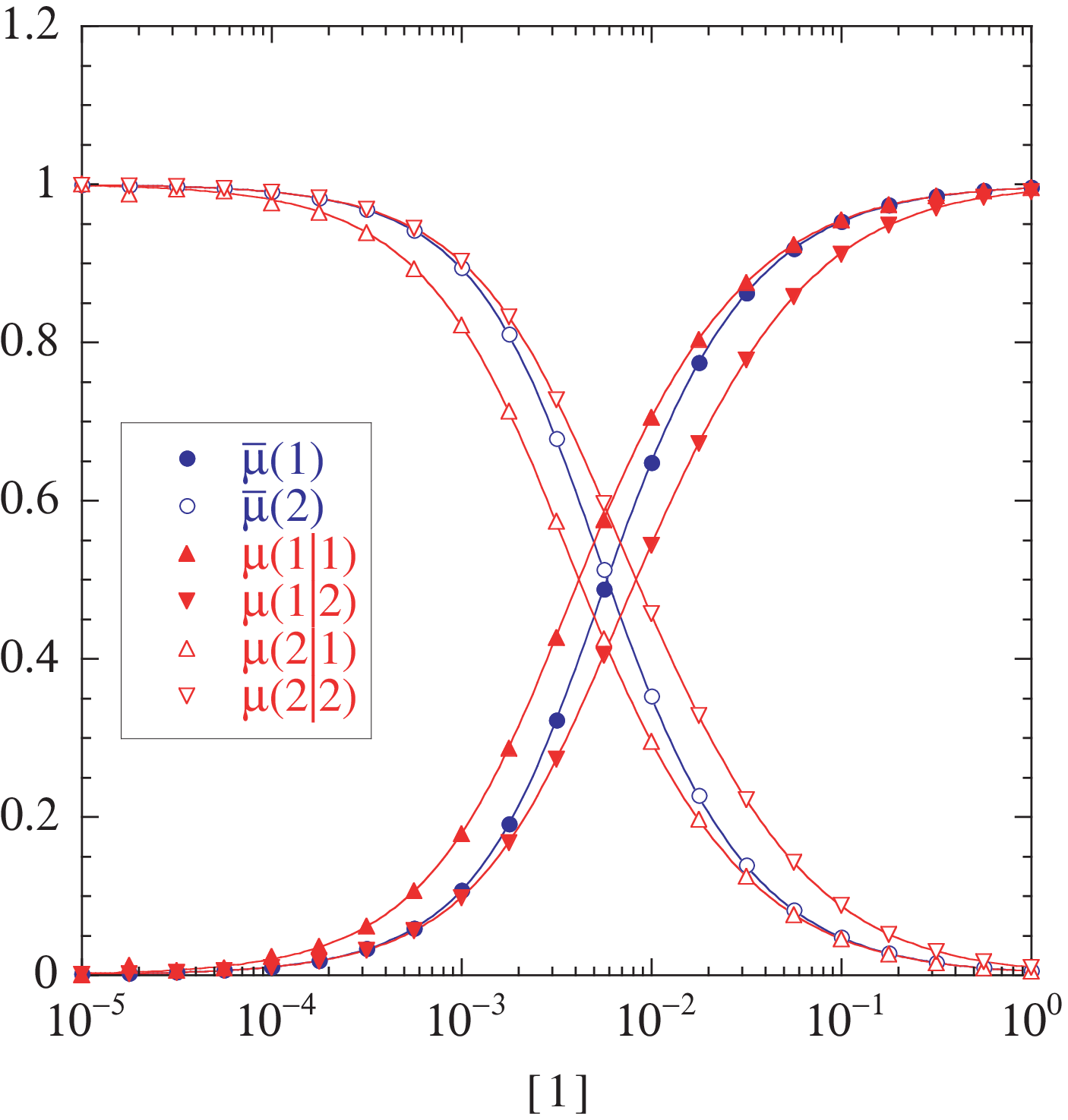}}}
\caption{Growth of a copolymer in the conditions (\ref{model1}): the bulk probabilities $\bar\mu(m)$ and the conditional probabilities $\mu(m\vert n)$ versus the concentration $[1]$ underlying the quantities plotted in Figs.~\ref{fig2} and~\ref{fig3}.  The dots are the results of simulations with Gillespie's algorithm, while the solid lines are the theoretical quantities obtained with Eqs.~(\ref{cond_proba_2}) and (\ref{bulk_proba}).}
\label{fig4}
\end{figure}

\begin{figure}[h]
\centerline{\scalebox{0.47}{\includegraphics{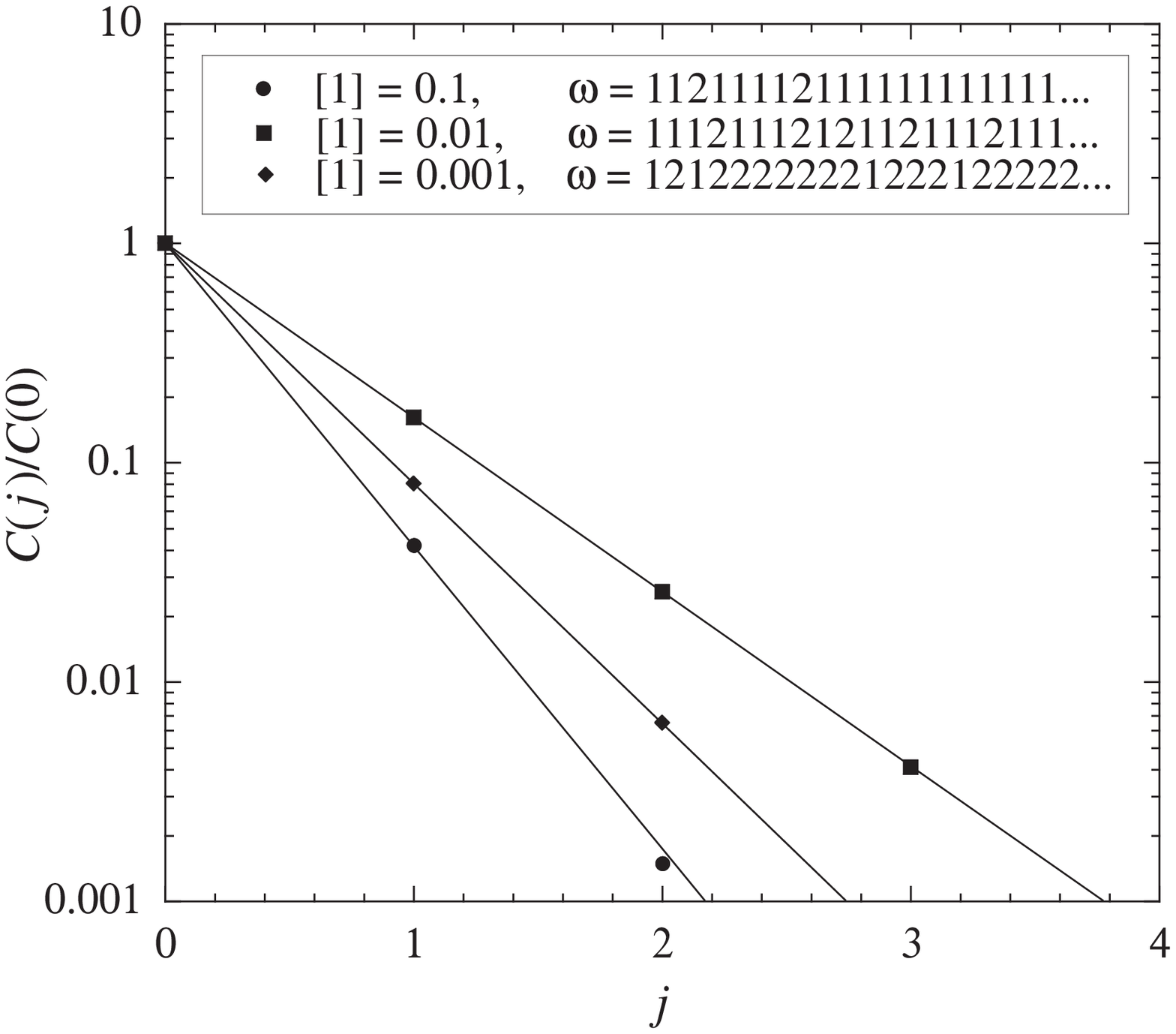}}}
\caption{Growth of a copolymer in the conditions (\ref{model1}): the normalized correlation function (\ref{correl-num}) versus the distance $j$ between successive monomeric units in the bulk of the copolymer sequence for three different concentrations $[1]$.  An example $\omega$ of copolymer sequence used in the sampling is shown in each case.  Averaging is done over $N=10^4$ sequences of long enough length $L+j_{\rm max}\gg 1$. The dots are the results of numerical simulations and the solid lines the theoretically expected decay (\ref{correl-theory-M=2}) with $\Lambda_2=0.0417$ if $[1]=0.1$, $\Lambda_2=0.1607$ if $[1]=0.01$, and $\Lambda_2=0.0806$ if $[1]=0.001$.}
\label{fig5}
\end{figure}

In Fig.~\ref{fig1}, the growth of the copolymer has been simulated over a time interval $t=200000$ and the length probability $p_t(l)$ defined by Eq.~(\ref{length_prob}) is plotted versus the length $l$, together with the probabilities $p_t(m;l)$ that are defined similarly, although by omitting the sum over the last monomeric unit $m=m_l$.   The results of the numerical simulations (dots) are compared with the following theoretical predictions (solid lines):
\be
p_t(l)\simeq \frac{1}{\sqrt{4\pi\, {\cal D} \, t}}\, \exp\left[ -\frac{(l-v\, t)^2}{4\, {\cal D}\, t}\right] ,
\label{ptl-diff}
\ee
and
\be
p_t(m;l) = \mu(m)\times p_t(l) \qquad \mbox{for}\quad m=1,2 \, ,
\label{ptml}
\ee
where the tip probabilities $\mu(m)$ are given by Eqs.~(\ref{tip_proba_2}), the mean growth velocity $v$ by Eq.~(\ref{velo2}), and the diffusivity by Eq.~(\ref{dispersion2}).  
We observe an excellent agreement between simulations and theory.  
As time increases, the probabilities~(\ref{ptl-diff}) and~(\ref{ptml}) become distributed over larger and larger copolymer lengths, which explains that the copolymer has the time to settle in the regime of steady growth where the probabilities take their self-consistent values predicted by theory.

Figure~\ref{fig2} depicts the growth velocity $v$, the free-energy driving force $\epsilon$, the Shannon disorder per monomer $D$, the affinity $A=\epsilon+D$, and the thermodynamic entropy production $d_{\rm i}S/dt=k_{\rm B} \, A\, v$, as a function of the concentration~$[1]$.  We see that the free-energy driving force is negative if~$[1]<[1]_{\rm c}=0.00654$, although the affinity and the velocity remain positive. Indeed, in this regime the entropic effect of the Shannon disorder in the sequence is large enough to drive the growth.\cite{AG08}  The free-energy driving force becomes positive if~$[1]>[1]_{\rm c}=0.00654$. 

Figure~\ref{fig3} depicts the same quantities in a log-log plot,  revealing that the velocity, the disorder, and the affinity here vanish proportionally to the concentration $[1]$.  Accordingly, the entropy production vanishes as $[1]^2$.  We also observe that the Shannon disorder reaches a maximum value $D\simeq 0.6783$ close to $[1]=[1]_{\rm d}\simeq 0.00596$.  

To understand this behavior, we plot in Fig.~\ref{fig4} the bulk probabilities~(\ref{bulk_proba}) and the conditional probabilities~(\ref{cond_proba_2}) as a function of the concentration $[1]$.  A crossover exists around the concentration $[1]_{\rm d}\simeq 0.00596$ between copolymers mainly composed of monomeric units $m=2$ if~$[1]<[1]_{\rm d}$ and copolymers with a majority of units $m=1$ if~$[1]>[1]_{\rm d}$.  At the crossover $[1]=[1]_{\rm d}$, the copolymers are random mixtures of both units so that the Shannon disorder per monomer is maximum.   

The conditional probabilities defining the first-order Markov chain control the statistical properties of the copolymer as well as the bulk probabilities (\ref{bulk_proba}).  Accordingly, the variations of the bulk probabilities follow those of the conditional probabilities.  Since $\mu(1\vert 1) \neq \mu(1\vert 2)$ and $\mu(2\vert 1) \neq \mu(2\vert 2)$, the chain is not Bernoullian, but the Markovian character is weak because these conditional probabilities are close to each other.

This is confirmed by examining the correlation functions (\ref{correl}) in Fig.~\ref{fig5}. These functions are computed as
\be
C(j) =  \frac{1}{N} \sum_{n=1}^N \left[ \frac{1}{L}\sum_{i=1}^L m_i^{(n)} m_{i+j}^{(n)} -\left(\frac{1}{L}\sum_{i=1}^L m_i^{(n)}\right)^2 \right] 
\label{correl-num}
\ee
with $0\le j\le j_{\rm max}$.  Moreover, the correlation functions are normalized to $C(j)/C(0)$.  We observe that the correlation functions rapidly decay as the distance $j$ between the monomeric units increases.  Therefore, the chains are close to being Bernoullian (in which case correlations would be equal to zero for $j\geq 1$).  The solid lines depict the theoretical expectation (\ref{correl-eigen}) of an exponential decay
\be
C(j)/C(0) = \left(\Lambda_2\right)^j \, ,
\label{correl-theory-M=2}
\ee
where $\vert\Lambda_2\vert <1$ is the second eigenvalue of the matrix of conditional probabilities given by solving Eqs.~(\ref{eigenvalue_problem}). 

If $[1]=0.01$, the differences seen in Fig.~\ref{fig4} between $\mu(m\vert 1)$ and $\mu(m\vert 2)$ for $m=1$ and $m=2$ are larger than if $[1]=0.1$ or $[1]=0.001$. Since these differences measure the departure from the Bernoullian character, the decay of the correlation function is slower for $[1]=0.01$ than for $[1]=0.1$ and $[1]=0.001$, as seen in Fig.~\ref{fig5}.

We notice that, according to Eq.~(\ref{mu_k-2}), the probability to find the monomeric unit $m$ at the $k^{\rm th}$ place behind the growing tip  converges to the bulk probability $\bar\mu(m)$ at the same rate $\Lambda_2$:
\be
\mu^{(-k)}(m) = \bar\mu(m)+ \left(\Lambda_2\right)^k \left[ \mu(m)-\bar\mu(m)\right] \, ,
\label{mu_k-M=2}
\ee
for $k=0,1,2,...$.

\subsection{Copolymerization and depolymerization with $M=2$ monomers}

Next, we consider the following values for the rate constants and the concentration of monomers $m=2$:
\bea
&& k_{+1\vert 1} = 0.1 \, , \quad\ k_{+1\vert 2} = 2 \, , \quad\ \ k_{+2\vert 1} = 3 \, , \qquad k_{+2\vert 2} =  0.4 \, ,\nonumber\\
&& k_{-1\vert 1} = 0.001\, , \  k_{-1\vert 2} = 0.02\, , \  k_{-2\vert 1} = 0.003\, , \ k_{-2\vert 2} = 0.04 \, , \nonumber\\
&& [2] = 0.005 \, .
\label{model2}
\eea
This set of values is more generic than the previous one in the sense that the equilibrium concentration of monomers $m=1$ is here non vanishing: 
\be
[1]_{\rm eq}=1.597\times 10^{-3} \, .
\label{model2_equil}
\ee
Copolymerization occurs if $[1]>[1]_{\rm eq}$ and depolymerization if $[1]<[1]_{\rm eq}$.

\subsubsection{Copolymerization}

\begin{figure}[h]
\centerline{\scalebox{0.5}{\includegraphics{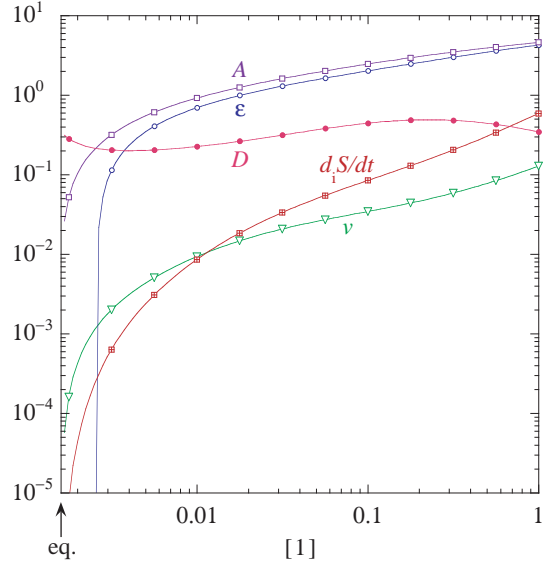}}}
\caption{Growth of a copolymer in the conditions (\ref{model2}) above the equilibrium concentration (\ref{model2_equil}):  The growth velocity $v$, the free-energy driving force $\epsilon$, the Shannon disorder per monomer $D$, the affinity $A=\epsilon+D$, and the thermodynamic entropy production $d_{\rm i}S/dt=A\, v$ with $k_{\rm B}=1$ versus the concentration $[1]$ of monomeric units $m=1$.  The dots are the results of simulations with Gillespie's algorithm, while the solid lines are the theoretical quantities obtained with Eqs.~(\ref{cond_proba_2})-(\ref{velo2}) and~(\ref{entrprod})-(\ref{disorder}).  The equilibrium concentration (\ref{model2_equil}) is marked by the vertical arrow.}
\label{fig6}
\end{figure}

\begin{figure}[h]
\centerline{\scalebox{0.5}{\includegraphics{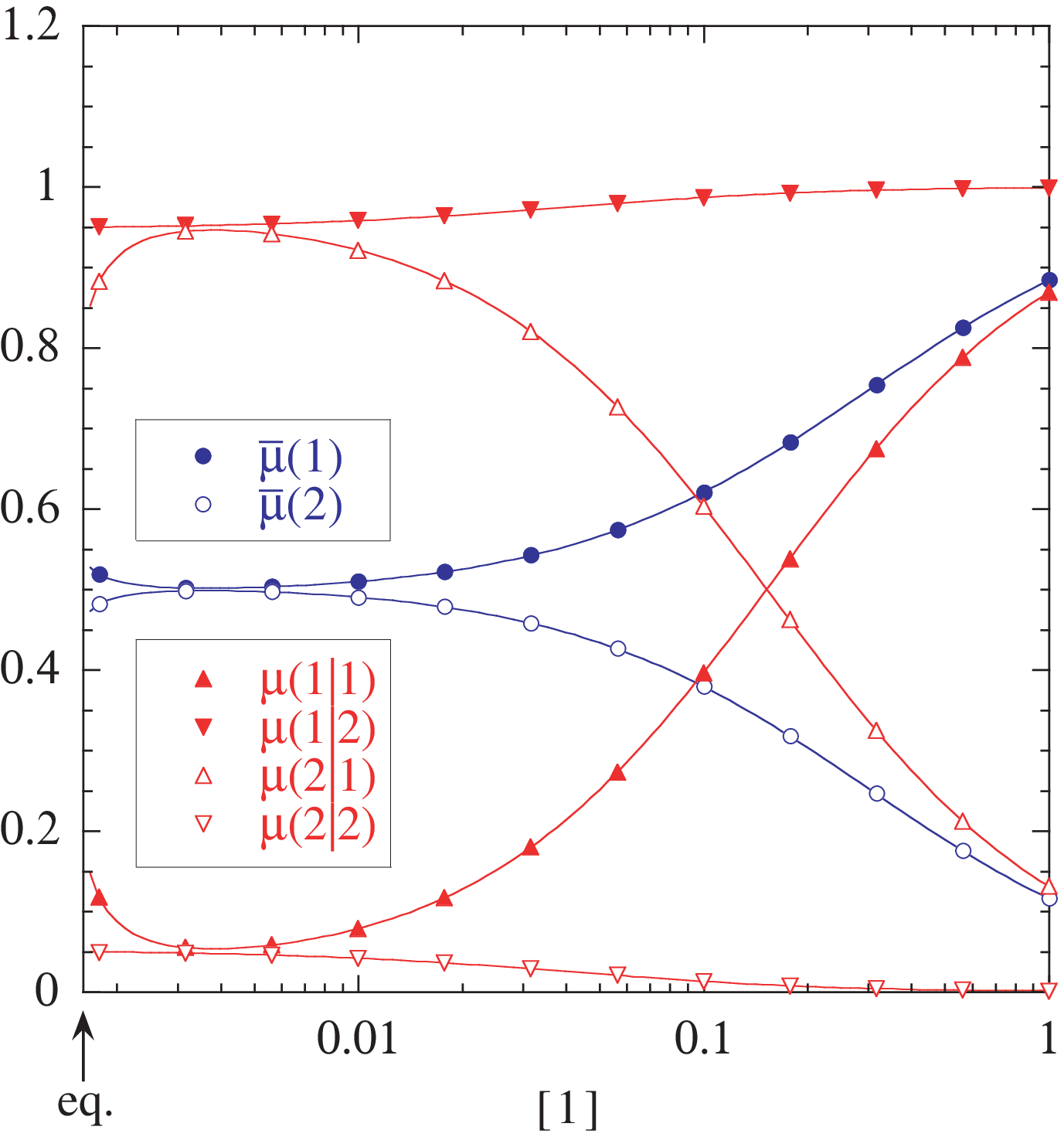}}}
\caption{Growth of a copolymer in the conditions (\ref{model2}) above the equilibrium concentration (\ref{model2_equil}): the bulk probabilities $\bar\mu(m)$ and the conditional probabilities $\mu(m\vert n)$ versus the concentration $[1]$ underlying the quantities plotted in Fig.~\ref{fig6}.  The dots are the results of simulations with Gillespie's algorithm, while the solid lines are the theoretical quantities obtained with Eqs.~(\ref{cond_proba_2}) and (\ref{bulk_proba}). The equilibrium concentration (\ref{model2_equil}) is marked by the vertical arrow.}
\label{fig7}
\end{figure}

\begin{figure}[h]
\centerline{\scalebox{0.47}{\includegraphics{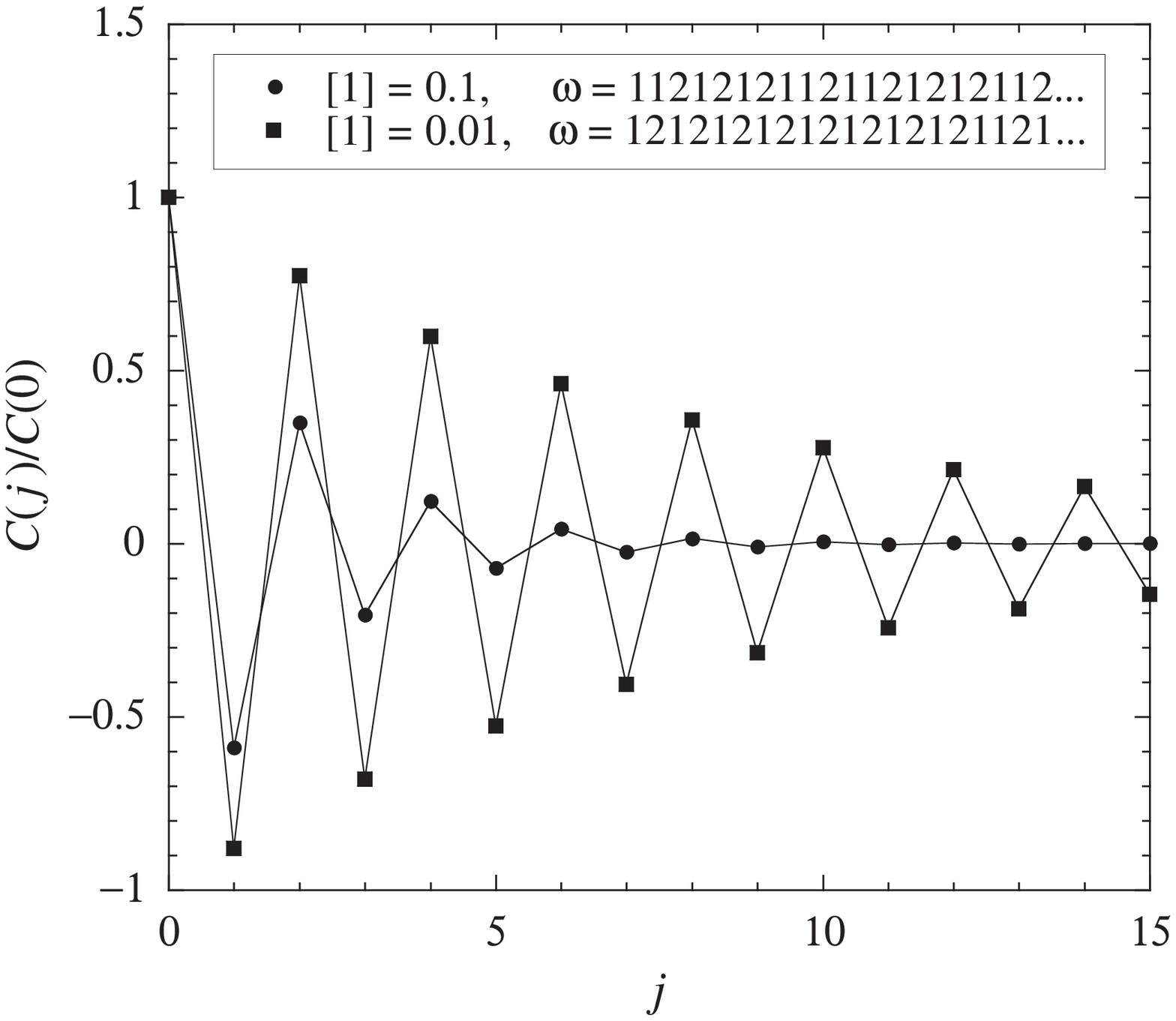}}}
\caption{Growth of a copolymer in the conditions (\ref{model2}) above the equilibrium concentration (\ref{model2_equil}): the normalized correlation function (\ref{correl-num}) versus the distance $j$ between successive monomeric units in the bulk of the copolymer sequence for two different concentrations $[1]$.  An example $\omega$ of copolymer sequence used in the sampling is shown in each case.  The dots are the results of numerical simulations and the solid lines the theoretically expected decay (\ref{correl-theory-M=2}) with $\Lambda_2=-0.5901$ if $[1]=0.1$, and $\Lambda_2=-0.8795$ if $[1]=0.01$.}
\label{fig8}
\end{figure}

Figure~\ref{fig6} depicts the mean growth velocity $v$, the free-energy driving force $\epsilon$, the Shannon disorder per monomer $D$, the affinity $A=\epsilon+D$, and the entropy production (\ref{entrprod}) obtained by numerical simulations (dots) and theory (solid lines) in the regime of copolymerization.  As expected, the velocity, the affinity, and the entropy production vanish at the equilibrium concentration (\ref{model2_equil}) where Eq.~(\ref{eps-D-equil}) holds with $D_{\rm eq}=0.326$.

The transition between the growth regimes by the entropic effect of sequence disorder and by free-energy driving occurs at the critical concentration $[1]_{\rm c}=0.00256$ where $\epsilon=0$ and the entropy production per monomer takes the value of the Shannon disorder in units of Boltzmann's constant: $(k_{\rm B} v)^{-1}d_{\rm i}S/dt=D=0.215$.

In the range of concentrations $[1]$ shown in Fig.~\ref{fig6}, the Shannon disorder is maximum around $[1]_{\rm d}=0.230$ where $D=0.490$. As in the previous example, this behavior can be understood in terms of the conditional and bulk probabilities plotted in Fig.~\ref{fig7}.  A first observation is that the bulk concentrations are non-monotonous functions of the concentrations of monomers.  Above $[1]_{\rm d}$, the bulk probability $\bar\mu(1)$ increases so that the copolymer becomes composed of a majority of monomeric units $m=1$.  Below $[1]_{\rm d}$, the bulk probabilities approach the values $\bar\mu(1)=\bar\mu(2)=0.5$ while the conditional probabilities are close to the values $\mu(1\vert 2)=\mu(2\vert 1)=1$ and $\mu(1\vert 1)=\mu(2\vert 2)=0$, which correspond to the formation of the alternating copolymer $12121212\cdots$.  In between, the copolymer sequence presents a random mixture of the periodic sequences $12121212\cdots$ and $11111111\cdots$, while the disorder is maximum at $[1]_{\rm d}=0.230$.

This is confirmed in Fig.~\ref{fig8} depicting the correlation functions~(\ref{correl-num}) at two different concentrations $[1]$.  At $[1]=0.01$, the correlation function presents slowly damped oscillations characteristic of an alternating copolymer sequence with a small amount of disorder.  The behavior is described by Eq.~(\ref{correl-theory-M=2}) with a negative eigenvalue close to the value $\Lambda_2=-1$, which it would have for the periodic sequence $12121212\cdots$.  At $[1]=0.1$, the disorder is stronger so that the correlation function decays faster, although the alternating character of the sequence is still present because $\Lambda_2$ remains negative.  

\subsubsection{Depolymerization}

\begin{figure}[h]
\centerline{\scalebox{0.475}{\includegraphics{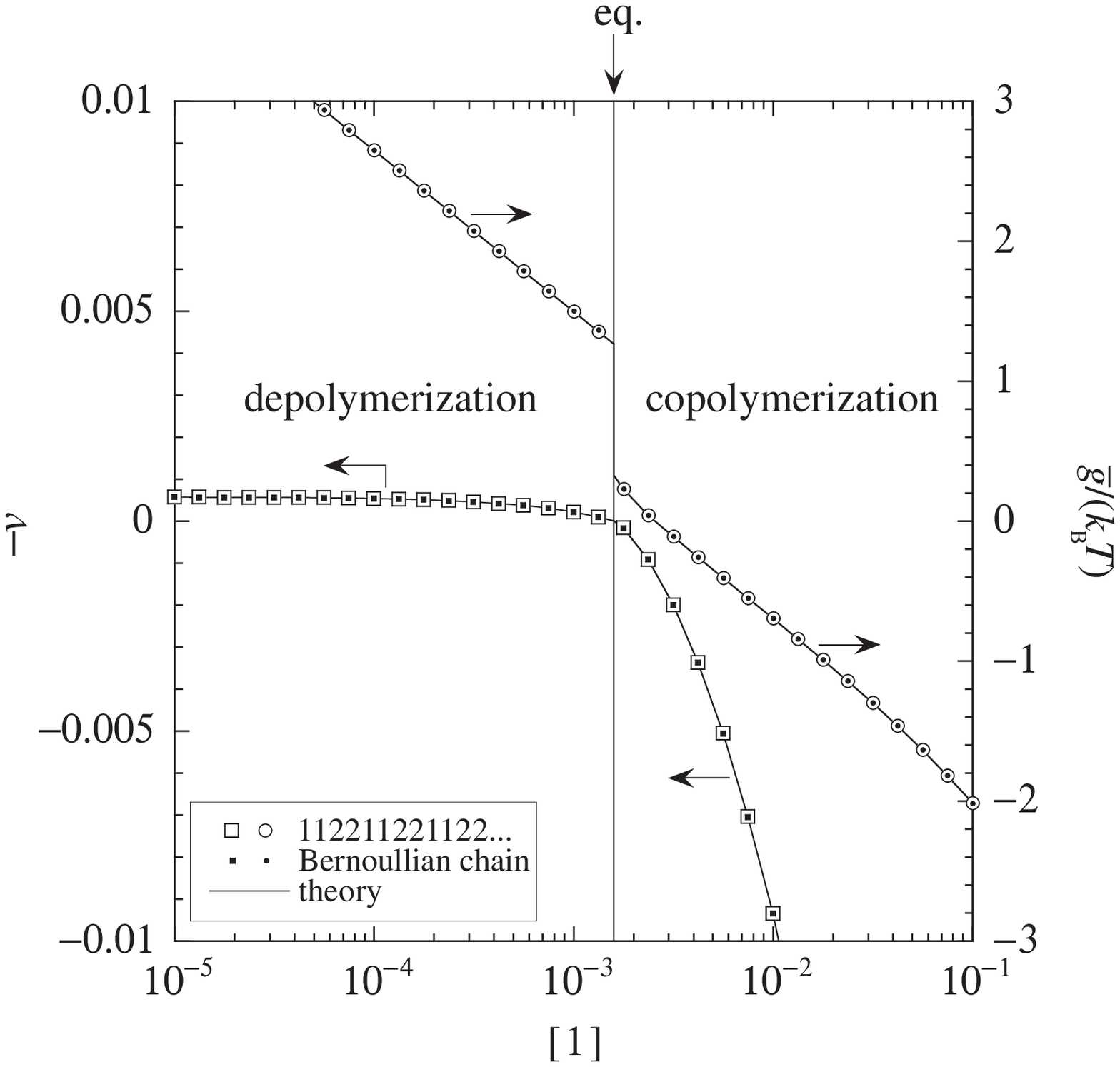}}}
\caption{Depolymerization and copolymerization of a chain in the conditions (\ref{model2}) below and above the equilibrium concentration~(\ref{model2_equil}): the depolymerization velocity $-v$ (open and filled squares, vertical axis on the left-hand side) and the free enthalpy per monomer $\bar g/(k_{\rm B}T)$ (open and filled circles, vertical axis on the right-hand side) versus the concentration $[1]$ of monomers $m=1$.  The theoretical predictions of Eqs.~(\ref{velo2})-(\ref{eps}) in the copolymerization regime and Eqs.~(\ref{mean_velo})-(\ref{eps-depolym}) in the depolymerization regime are shown by solid lines.  The open squares and circles are the results for an initial periodic sequence $11221122\cdots$, and the filled squares and circles the results for an initial Bernoullian sequence with probabilities $\bar\mu_1(1)=\bar\mu_1(2)= 0.5$.  The regimes of depolymerization with $v<0$ and copolymerization with $v>0$ are separated by the vertical line located at the equilibrium concentration (\ref{model2_equil}).}
\label{fig9}
\end{figure}

\begin{figure}[h]
\centerline{\scalebox{0.45}{\includegraphics{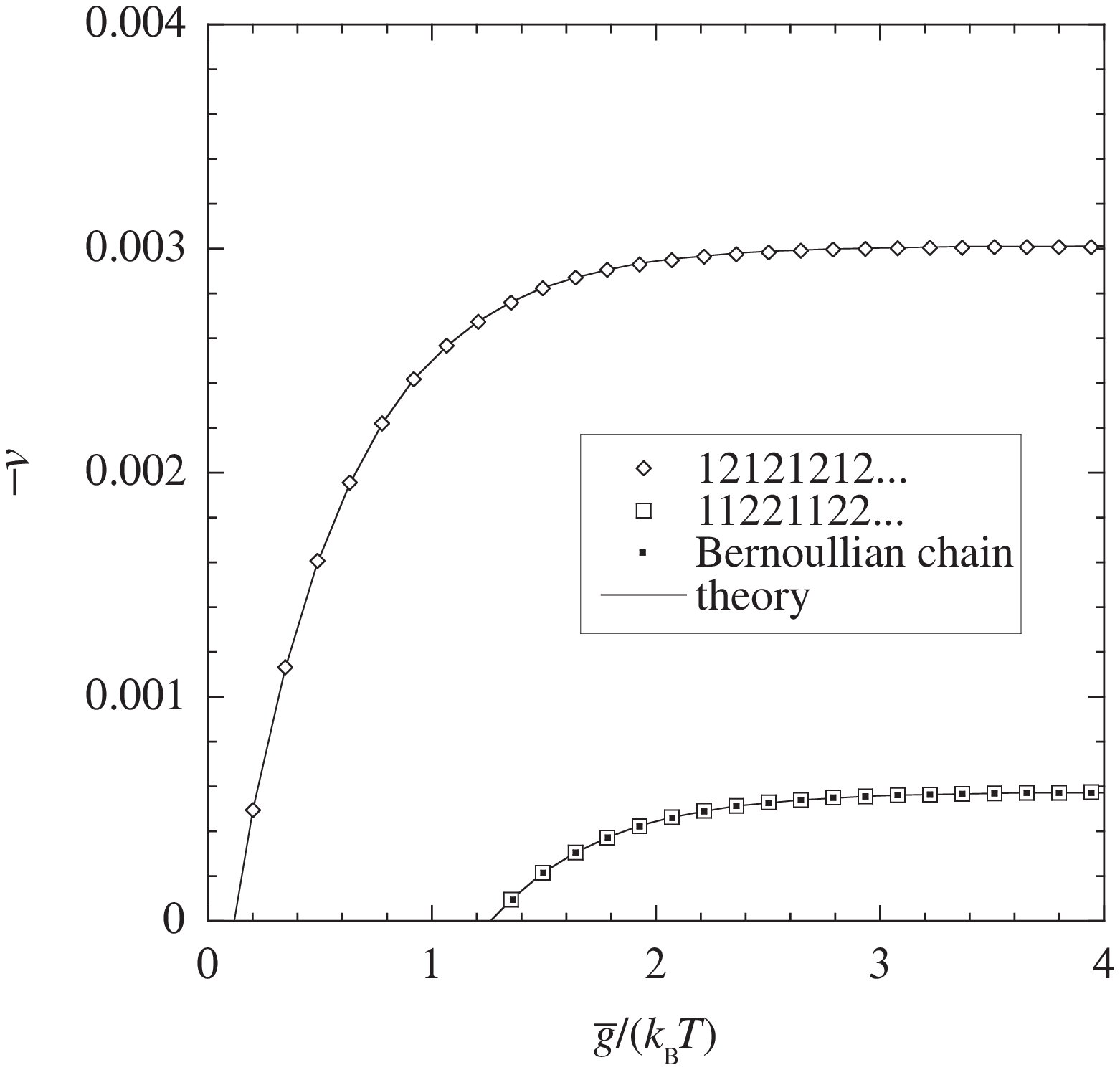}}}
\caption{Depolymerization of a chain in the conditions (\ref{model2}) below the equilibrium concentration (\ref{model2_equil}): the depolymerization velocity $-v$ versus the free enthalpy per monomer $\bar g/(k_{\rm B}T)$ for the periodic sequences $12121212\cdots$ (open diamonds) and $11221122\cdots$ (open squares), and a Bernoullian sequence with probabilities $\bar\mu_1(1)=\bar\mu_1(2)= 0.5$ (filled squares).   The solid lines show the theoretical predictions given by Eqs.~(\ref{mean_velo})-(\ref{eps-depolym}).}
\label{fig10}
\end{figure}

In order to study the regime of depolymerization, kinetic Monte Carlo simulations are carried out starting from an initial copolymer sequence generated either as a Bernoullian chain with probabilities $\bar\mu_1(1)=\bar\mu_1(2)=0.5$, or as the periodic chains $12121212\cdots$ and $11221122\cdots$.  Figure~\ref{fig9} shows the mean velocity of the tip and the free enthalpy per monomer defined by Eqs.~(\ref{eps}) or~(\ref{eps-depolym}).  The mean velocity is positive in the copolymerization regime $[1]>[1]_{\rm eq}$ and negative in the depolymerization regime $[1]<[1]_{\rm eq}$.  The theoretical predictions are different in both regimes.  During copolymerization, the sequence is self generated by the process and the mean velocity is given by Eqs.~(\ref{central_eqs})-(\ref{velo2}) in terms of the self-consistent conditional and tip probabilities (\ref{cond_proba_2})-(\ref{tip_proba_2}).  In contrast, during depolymerization, the sequence is generated {\it a priori} and the mean velocity is determined by Eq.~(\ref{mean_velo}) in terms of the dyad distribution $\bar\mu_2(mn)$ in the initial sequence.
Both theoretical curves for the velocity vanish at equilibrium in Fig.~\ref{fig9}.  However, they have different slopes at $v=0$ because the mechanism of front propagation is different between the two regimes.

The free enthalpy is also different between the copolymerization and depolymerization regimes.  At equilibrium, this quantity presents a discontinuity in Fig.~\ref{fig9}, which can be explained as follows.  During copolymerization, since the copolymer sequence is self generated, the equilibrium value of the free enthalpy is related to the Shannon disorder per monomer by Eq.~(\ref{eps-D-equil}) with $D_{\rm eq}=0.326$.  In contrast, during depolymerization, the sequence has already been generated under different conditions prior to the process.  For the Bernoullian sequence, the quantities characterizing disorder are $\bar I_2=\bar I_\infty = \ln 2$.  For the periodic sequence $11221122\cdots$, they are $\bar I_2= \ln 2$ and $\bar I_\infty = 0$.  For both sequences, $\bar I_2= \ln 2$ but the free enthalpy of depolymerization is equal to $\bar g= 1.264\times k_{\rm B}T$.  Hence, a discontinuity in $\bar g/(k_{\rm B}T)$ occurs, as seen in Fig.~\ref{fig9}.

A remarkable feature is that the depolymerization process does not depend on statistical correlations in the initial sequence beyond dyads according to Eqs.~(\ref{mean_velo}) and (\ref{eps-depolym}).  In order to test this prediction, we compare in Fig.~\ref{fig9} the depolymerization of a Bernoullian sequence with the probabilities $\bar\mu_2(mn)=\bar\mu_1(m)\,\bar\mu_1(n)$ and $\bar\mu_1(1)=\bar\mu_1(2)=0.5$, to the depolymerization of the periodic chain $11221122\cdots$, which has the same probabilities $\bar\mu_2(mn)$ but different higher-order statistical correlations.  According to Eqs.~(\ref{mean_velo}) and (\ref{eps-depolym}), both chains should have the same depolymerization velocity and free enthalpy per monomer.  This prediction is confirmed in Fig.~\ref{fig9} by the coincidence of the open and filled dots in the depolymerization regime.  

Interestingly, Landauer's principle is satisfied for the present first-order Markovian processes.\cite{AG13} 
According to this principle, the erasure of information during depolymerization should dissipate a free energy larger than the Shannon information $\bar I_\infty$ per monomer counted in units of thermal energy $k_{\rm B}T$, as expressed by Eq.~(\ref{Binfty}).  Moreover, since the attachment and detachment rates only depend on the last monomeric unit at the tip of the copolymer, the bound is given by Eq.~(\ref{B2}) in terms of the Shannon information $\bar I_2$ in the dyads of the sequence, which is a stronger bound because $\bar I_2 > \bar I_\infty$.  In order to test this prediction, Fig.~\ref{fig10} depicts the mean depolymerization velocity $-v$ versus the dissipated free enthalpy per monomer in units of thermal energy, $\bar g/(k_{\rm B}T)$, for the depolymerization of the aforementioned Bernoullian and periodic sequences.  

For the periodic sequence $12121212\cdots$, the Shannon information is vanishing, $\bar I_2 = \bar I_\infty = 0$, and the velocity vanishes at the equilibrium concentration (\ref{model2_equil}) with the free enthalpy $\bar g=0.1126\times k_{\rm B}T$, which indeed satisfies the inequalities~(\ref{Binfty}) and~(\ref{B2}).  

For both the Bernoullian sequence and the periodic sequence $11221122\cdots$ characterized by the same dyad information $\bar I_2=\ln 2=0.693$, the velocity vanishes at the free enthalpy $\bar g=1.264 \times k_{\rm B}T$, again in agreement with the lower bounds~(\ref{Binfty}) and~(\ref{B2}).  The bound (\ref{B2}) is not reached because the conditions $\bar\mu_2(mn)=z_{n\vert m} \, \mu_{\rm eq}(m)$ required to meet this bound [and {\it a fortiori} the bound (\ref{Binfty})] are not satisfied for the rates (\ref{model2})-(\ref{model2_equil}) of the process with respect to the dyad distribution $\bar\mu_2(mn)$ of the initial sequences.  The bound (\ref{B2}) would only be met for the initial copolymer with a dyad distribution $\bar\mu_2(mn)$ matching the equilibrium distribution~(\ref{equil_cond}) determined by the rates according to $\mu_{\rm eq}(mn)=z_{n\vert m} \, \mu_{\rm eq}(m)$ with $z_{n\vert m}=w_{+n\vert m}/w_{-n\vert m}$ as discussed in Subsection~\ref{VC}.

\subsection{Copolymerization with $M=3$ monomers}

We continue with a copolymerization involving $M=3$ types of monomers, a so-called terpolymerization process.  The rate constants and the concentrations of monomers $m=2$ and $m=3$ are chosen as
\bea
&& k_{+1\vert 1} = 0.1 \, , \quad\ k_{+1\vert 2} = 2 \, , \quad\ \ \ k_{+2\vert 1} = 3 \, , \qquad k_{+2\vert 2} =  0.4 \, ,\nonumber\\
&& k_{-1\vert 1} = 0.001\, , \  k_{-1\vert 2} = 0.02\, , \ \ k_{-2\vert 1} = 0.003\, , \ k_{-2\vert 2} = 0.04  \, , \nonumber\\
&& k_{+1\vert 3} = 5 \, , \qquad  k_{+2\vert 3} = 0.005 \, , \ k_{+3\vert 1} = 0.1 \, , 
\quad\ k_{+3\vert 2} = 1 \, , \qquad k_{+3\vert 3} = 2 \, , \nonumber\\
&& k_{-1\vert 3} = 0.01 \, , \ \ k_{-2\vert 3} = 0.01 \, , \ \ \, k_{-3\vert 1} = 0.03  \, , \ \ \, k_{-3\vert 2} = 0.001 \, , \ k_{-3\vert 3} = 0.05  \, , \nonumber\\
&& [2] = 0.005 \, , \quad\ [3] = 0.01 \, .
\label{model3}
\eea
The equilibrium condition (\ref{Z-eq}) is satisfied if the concentration of monomers $m=1$ is equal to
\be
[1]_{\rm eq}=2.148\times 10^{-5}\, .
\label{model3_equil}
\ee

\begin{figure}[h]
\centerline{\scalebox{0.5}{\includegraphics{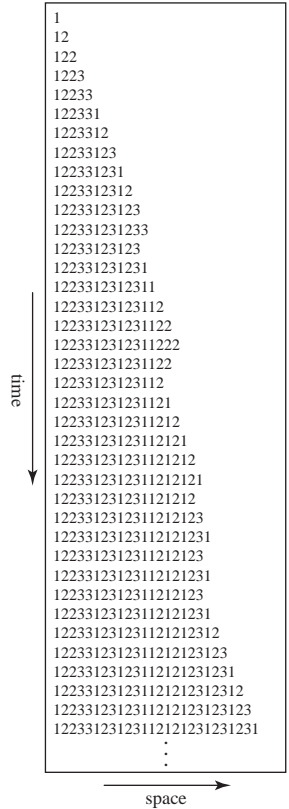}}}
\caption{Schematic space-time plot of the growth of a terpolymer in the conditions (\ref{model3}) at the concentration $[1]=0.01$.}
\label{fig11}
\end{figure}

\begin{figure}[h]
\centerline{\scalebox{0.5}{\includegraphics{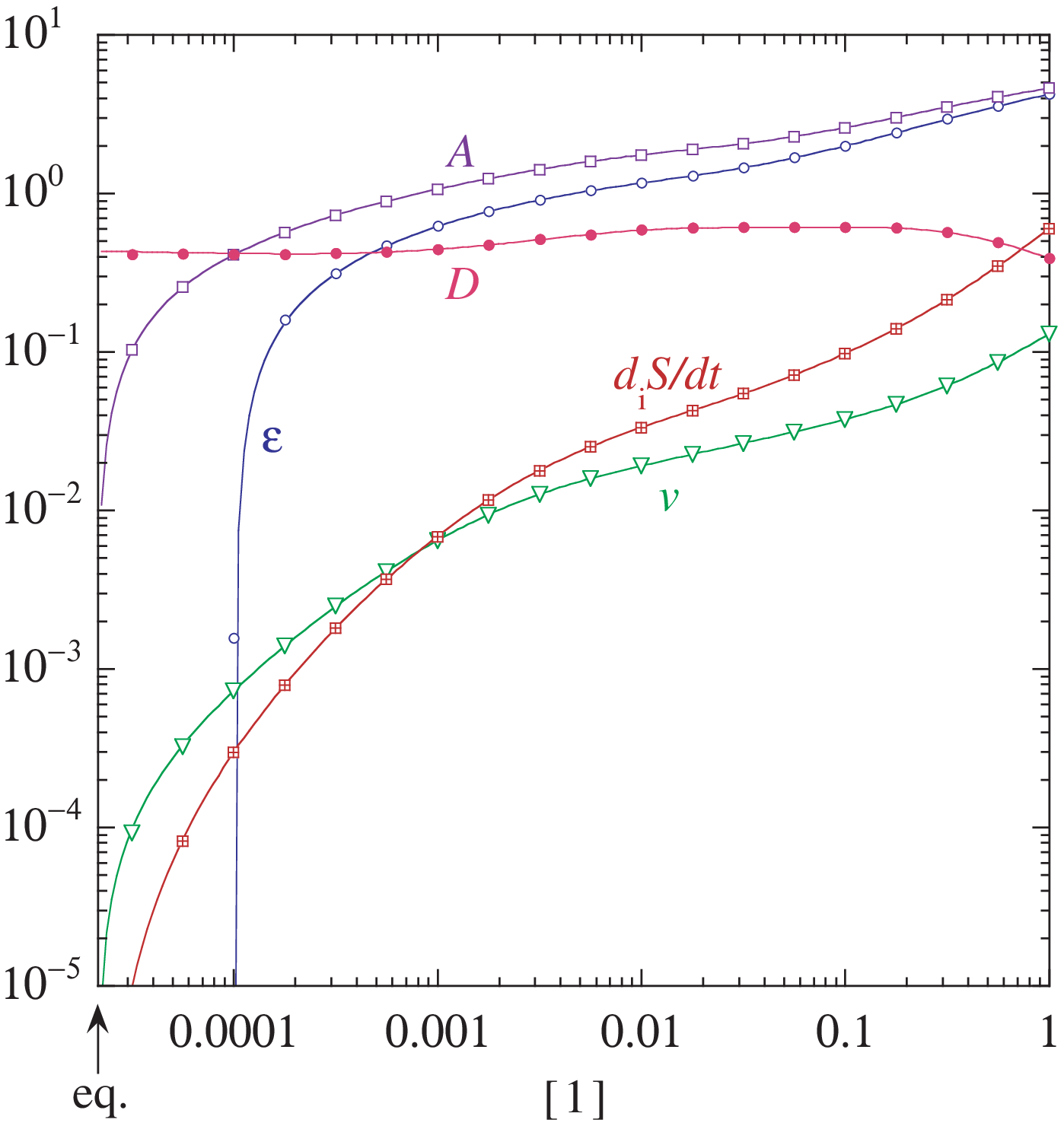}}}
\caption{Growth of a terpolymer in the conditions (\ref{model3}) above the equilibrium concentration (\ref{model3_equil}):  The growth velocity $v$, the free-energy driving force $\epsilon$, the Shannon disorder per monomer $D$, the affinity $A=\epsilon+D$, and the thermodynamic entropy production $d_{\rm i}S/dt=A\, v$ with $k_{\rm B}=1$ versus the concentration $[1]$ of monomeric units $m=1$.  The dots are the results of simulations with Gillespie's algorithm, while the solid lines are the theoretical quantities obtained with Eqs.~(\ref{cond_proba_2})-(\ref{velo2}) and~(\ref{entrprod})-(\ref{disorder}).  The equilibrium concentration (\ref{model3_equil}) is marked by the vertical arrow.}
\label{fig12}
\end{figure}

\begin{figure}[h]
\centerline{\scalebox{0.5}{\includegraphics{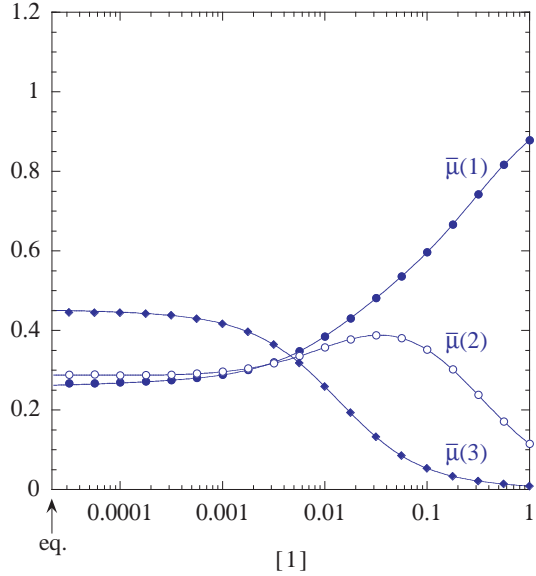}}}
\caption{Growth of a terpolymer in the conditions (\ref{model3}) above the equilibrium concentration (\ref{model3_equil}):  the bulk probabilities $\bar\mu(m)$ versus the concentration $[1]$.  The dots are the results of simulations with Gillespie's algorithm, while the solid lines are the theoretical quantities obtained with Eqs.~(\ref{bulk_proba}). The equilibrium concentration (\ref{model3_equil}) is marked by the vertical arrow.}
\label{fig13}
\end{figure}

\begin{figure}[h]
\centerline{\scalebox{0.45}{\includegraphics{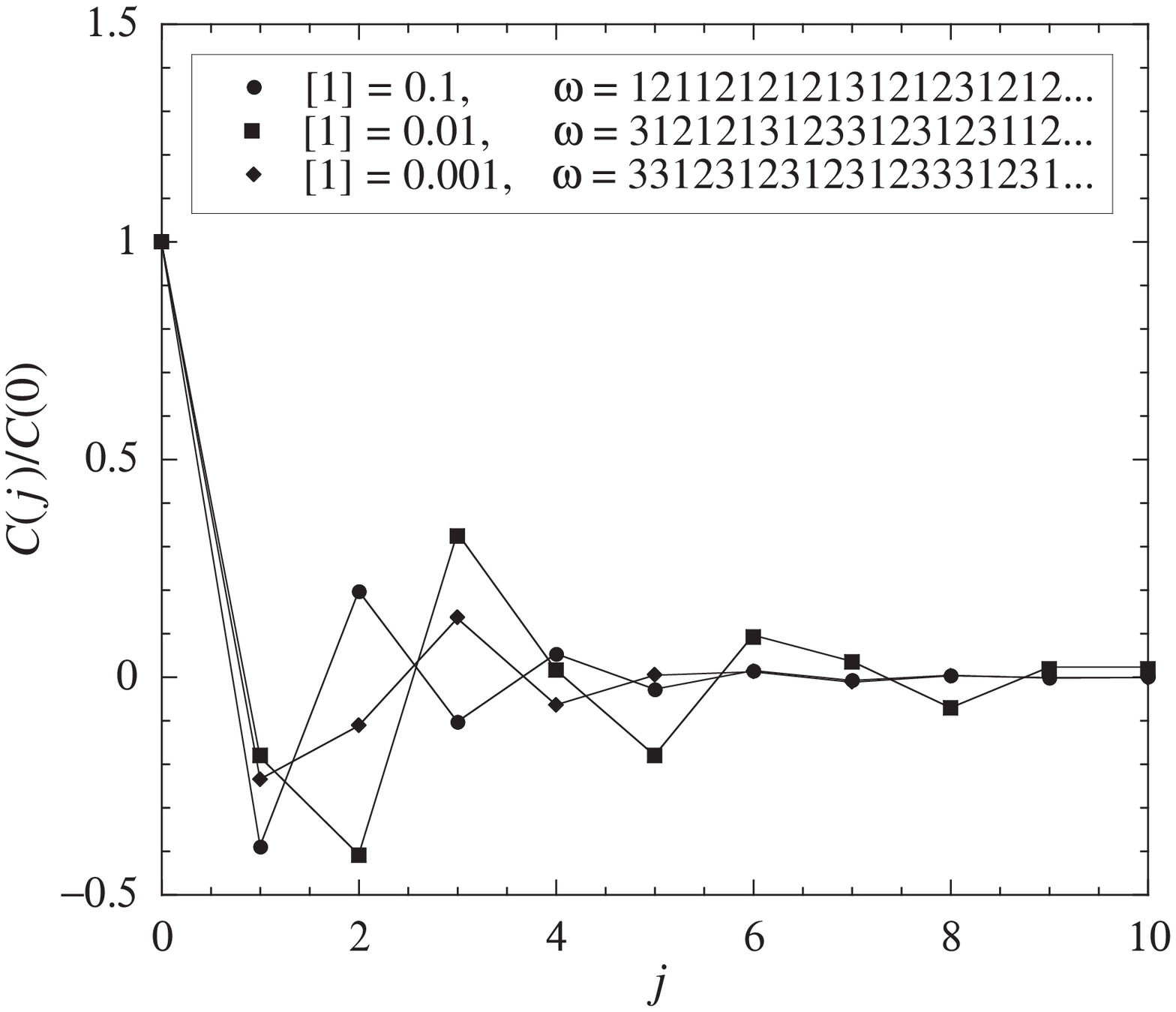}}}
\caption{Growth of a terpolymer in the conditions (\ref{model3}) above the equilibrium concentration (\ref{model3_equil}):  the normalized correlation function (\ref{correl-num}) versus the distance $j$ between successive monomeric units in the bulk of the terpolymer sequence for three different concentrations $[1]$ (filled circles if $[1]=0.1$; filled squares if $[1]=0.01$; filled diamonds if $[1]=0.001$).  An example $\omega$ of terpolymer sequence used in the sampling is shown in each case.  The dots are the results of numerical simulations and the solid lines the theoretically expected decay (\ref{correl-eigen}) with $\Lambda_2=-0.5205$ and $\Lambda_3=-0.0383$ if $[1]=0.1$, $\Lambda_{2,3}=0.5181\,\exp(\pm 2.279\, i)$ if $[1]=0.01$, and $\Lambda_{2,3}=0.7168\,\exp(\pm 1.989\, i)$ if $[1]=0.001$.}
\label{fig14}
\end{figure}

Figure~\ref{fig11} shows a schematic space-time plot of a growing terpolymer for $[1]=0.01$.  We see that the tip of the terpolymer grows at a positive mean velocity with fluctuations due to the stochasticity of the attachment and detachment events.

The dependences on the concentration $[1]$ of the mean velocity $v$, the free-energy driving force $\epsilon=-g/(k_{\rm B}T)$, the Shannon disorder per monomer $D$, the affinity $A=\epsilon+D$, and the thermodynamic entropy production $d_{\rm i}S/dt=k_{\rm B}\, v\, A$ are depicted in Fig.~\ref{fig12}.  As it should, the mean velocity, the affinity, and the entropy production are vanishing at the equilibrium concentration (\ref{model3_equil}) where $D_{\rm eq}=0.4315$.  The transition between the regime of disorder-driven growth where $\epsilon<0$ but $D>-\epsilon$ and the regime of free-energy driven growth where $\epsilon>0$ occurs at the critical concentration $[1]_{\rm c}=1.03\times 10^{-4}$ in this example.  The disorder reaches a maximum $D=0.6113$ at $[1]_{\rm d}=0.1061$.

The bulk probabilities of the monomeric units are plotted in Fig.~\ref{fig13} where we observe that the monomeric unit $m=3$ is more frequent for $[1]<0.00457$.  Around $[1]\simeq 0.00475$, the three monomeric units have nearly the same fraction $\bar\mu(1)\simeq\bar\mu(2)\simeq\bar\mu(3)\simeq 1/3$.  For  $[1]>0.18$, the fraction of the monomeric unit $m=1$ exceeds $\bar\mu(1)=2/3$ and the disorder decreases.  As in the previous example, the disorder is larger for sequences grown in the crossover region $0.00457<[1]<0.18$.

In Fig.~\ref{fig14}, the correlation function~(\ref{correl-num}) is shown for three different concentrations $[1]$.  We see that the correlation function takes positive and negative values at the three concentrations so that the terpolymers grown under these conditions have an alternating character.  According to Eq.~(\ref{correl-eigen}), the decay of the correlation function can be decomposed in terms of the eigenvalues of the matrix~(\ref{matrix-M}) of conditional probabilities.  Besides the eigenvalue $\Lambda_1=1$, the two other eigenvalues $\Lambda_2$ and $\Lambda_3$ are real and negative if $[1]=0.1$, but complex conjugate if $[1]=0.01$ and $[1]=0.001$.  For $[1]=0.1$, the fraction of monomeric units $m=3$ is small so that the alternating character is mainly between the monomeric units $m=1$ and $m=2$, which explains that the subleading eigenvalue $\Lambda_2=-0.5205$ is real and negative and mostly contributes to the correlation function seen in Fig.~\ref{fig14} (filled circles).  In contrast, if $[1]=0.01$ or $[1]=0.001$, the monomeric fraction is better distributed among the three units, allowing a triadic alternance in relation with the fact that the subleading eigenvalues are complex conjugate, $\Lambda_2=\Lambda_3^*$.

\section{Conclusions}
\label{Conclusions}

In the present paper, we have developed a theory for the kinetics and thermodynamics of living copolymerization and depolymerization processes with attachment and detachment rates depending on the last monomeric unit at the tip of the copolymer.  We have adopted a stochastic description based on coarse graining at the level of the sequence of monomeric units composing the copolymer. 

In this framework, we have shown that the kinetic equations can be solved in the regimes of steady growth or depolymerization.   The growing copolymer is described by a first-order Markov chain.  Analytical expressions are obtained for the conditional probabilities of this Markov chain, for the tip and bulk probabilities of the monomeric units, and for the mean growth velocity.  Moreover, the thermodynamic entropy production is given in terms of the growth velocity, the free-energy driving force, and the Shannon disorder in the Markov chain.  In the fully irreversible regime of copolymerization of $M=2$ monomers, the Mayo-Lewis equation is recovered for the ratio of mole fractions of the two monomeric units.\cite{ML44,AG44,FR99}

Expressions are also derived for related quantities in the regime of depolymerization.  However, depolymerization is different than copolymerization because it starts from an initial copolymer with arbitrary statistical properties, contrary to a growing copolymer which is self generated by its formation process.  In this regard, the thermodynamic entropy production of copolymerization has a contribution given by the Shannon disorder per monomer because of the exponential proliferation of different possible sequences for growing copolymers.  This contribution is absent for depolymerization since the statistical properties of the sequence remain those of the initial one.  However, during depolymerization, the entropy production obeys Landauer's principle, according to which the dissipated free energy cannot be smaller than the erased information in units of the ambient thermal energy.\cite{AG13}  Here, since the process only depends on the last monomeric unit of the copolymer, the lower bound is given by the Shannon information in the dyads composing the copolymer.

Thermodynamic equilibrium is found between the regimes of copolymerization and depolymerization.

In the limit where the attachment and detachment rates no longer depend on the last monomeric unit at the tip of the copolymer, the present theory reduces to the one we have established for Bernoullian chains in our previous work.\cite{AG09}

The theoretical predictions are in excellent agreement with the results of kinetic Monte Carlo simulations by Gillespie's algorithm.\cite{G76,G77}  The examples we studied confirm that the growth of first-order Markov chains can be driven by the disorder of the sequence in an adverse free-energy landscape.\cite{AG08,J08} This disorder-driven regime exists between thermodynamic equilibrium and the regime of growth in a favorable free-energy landscape.  The examples also show that the properties only depend on the dyad composition of the copolymer for the class of processes here considered.  Our simulations reveal complex, non-linear behaviors in copolymer self-organization. In particular, the composition and the disorder of a copolymer change non-monotonously as the concentrations of monomers are varied.

The present theory can be applied to obtain quantitative predictions for different living copolymerization reactions\cite{OCY97,YMIMSMMNTOYMF04,ZN05,MSSKO06,SMNK10,LZFWLZ13,WLHLZW13} under conditions where the detachment rates cannot be neglected and the available results\cite{ML44,AG44,FR99} would not apply.  The theory could be particularly useful for living ring-opening copolymerization.\cite{DPDMJ96,DK09,NP13}  More generally, it may contribute to the control of the composition of chains produced by living copolymerization, as well as other self-organization processes.\cite{WSH12}  In the Bernoullian case, we demonstrated with Eq.~(\ref{mstar}) that copolymers of defined composition can be grown arbitrarily close to or far from equilibrium.  

Our results also apply to biological copolymerization processes such as DNA replication where the attachment rates of nucleotide triphosphate depend on the previously incorporated nucleotide.\cite{J93}

Extensions to the copolymerization of higher-order Markov chains can be envisaged.  We hope to report on these issues in the future.

\begin{acknowledgments}
The authors thank Prof.~Yves~Geerts for helpful discussions.
This research is financially supported by the Universit\'e Libre de Bruxelles and the Belgian Federal Government under the Interuniversity Attraction Pole project P7/18 ``DYGEST".
\end{acknowledgments}


\appendix

\section{The case of Bernoullian chains}
\label{AppA}

Bernoullian chains are generated when the rates do not depend on the previous monomeric units:
\be
w_{\pm m_l\vert m_{l-1}} = w_{\pm m_l}\, .
\label{rates_B}
\ee
In this case, the thermodynamic equilibrium condition (\ref{NSC_eq}) reads
\be
\sum_{m=1}^M \frac{w_{+ m}}{w_{-m}} = 1 \, .
\ee
Indeed, $(1, 1, \ldots, 1)$ is a left eigenvector of the matrix ${\boldsymbol{\mathsf Z}}$ with eigenvalue $\sum_{m} w_{+ m}/w_{-m}$. Since this vector is positive, its corresponding eigenvalue is the largest eigenvalue of ${\boldsymbol{\mathsf Z}}$ by the Perron-Frobenius theorem.\cite{M00}

Copolymerization manifests itself if $\sum_{m} w_{+ m}/w_{-m} > 1$. 
For the rates (\ref{rates_B}), Eqs.~(\ref{central_eqs}) show that all the partial velocities take the same value, $v_m=v$, which satisfies
\be
\sum_{m=1}^M \frac{w_{+ m}}{w_{-m} + v} = 1 \, .
\label{velo_B}
\ee
We notice that, for given rates $w_{\pm m}$, the solution of this equation is unique as its left-hand side is a decreasing function of $v$. 

Equations (\ref{tip_proba_2}) give the probabilities
\be
\mu(m) = \frac{w_{+ m}}{w_{-m} + v} 
\label{mu_B}
\ee
for $m=1,2,...,M$, and the conditional probabilities~(\ref{cond_proba_2}) become equal to these probabilities
\be
\mu(m\vert n) = \mu(m)
\ee
for $m,n=1,2,...,M$.  Consequently, the Markov chain reduces to a Bernoullian chain with the probabilities $\{\mu(m)\}_{m=1}^M$.  
The mean growth velocity is given by the positive root of Eq.~(\ref{velo_B}), and thus 
\be
v = \sum_{m=1}^M \left[w_{+m} - w_{-m} \, \mu(m)\right] \, .
\ee
For a Bernoullian chain, the tip probabilities coincide with the bulk probabilities, $\mu(m)=\bar\mu(m)$, so that there is no need to make a distinction between them in this case.

During copolymerization, the thermodynamic entropy production is given by Eq.~(\ref{entrprod}), but with the free-energy driving force
\be
\epsilon = -\frac{g}{k_{\rm B}T} = \sum_{m=1}^M \mu(m) \, \ln \frac{w_{+m}}{w_{-m}}
\label{eps_B}
\ee
and the Shannon disorder per monomer
\be
D = - \sum_{m=1}^M \mu(m) \, \ln \mu(m) \, ,
\label{disorder_B}
\ee
as expected for a Bernoullian chain.
The results of Ref.~\onlinecite{AG09} are thus recovered.

Depolymerization occurs if $\sum_{m} w_{+ m}/w_{-m} < 1$.  For processes with the rates~(\ref{rates_B}), the mean velocity~(\ref{mean_velo}) reduces to the formula
\be
v = - \frac{ 1 - \sum_{m=1}^M w_{+m}/w_{-m} }{ \sum_{m=1}^M \bar\mu_1(m)/w_{-m} } \, ,
\ee
as given in Ref.~\onlinecite{AG13}.\\

Notably, the self-organization of Bernoullian copolymers can be fully controlled  in order to generate a structure of defined composition. 

A copolymer with composition $\{\mu(m)\}_{m=1}^M$ can be synthesized at the arbitrary speed $v > 0$ by choosing the concentrations
\bea
[m]^{*} = \mu(m) \, \frac{k_{-m}}{k_{+m}} \paren{1+\frac{v}{k_{-m}}} \, .
\label{mstar}
\eea
Indeed, these concentrations satisfy Eqs.~(\ref{velo_B}) and~(\ref{mu_B}). 

The corresponding entropy production is given by Eq.~(\ref{entrprod}) with
\bea
A = \sum_{m=1}^M \mu(m) \ln \paren{1+\frac{v}{k_{-m}}} \, . 
\label{A_control}
\eea
The affinity is an increasing function of the speed $v$ for a given composition. 

A copolymer of defined composition can thus be grown arbitrarily close to or far from equilibrium.
This result can have important implications for our understanding of the self-organization of multicomponent structures.



\begin{thebibliography}{99}

\bibitem{ML44} F.~R.~Mayo and F.~M.~Lewis, J. Am. Chem. Soc. {\bf 66}, 1594 (1944).

\bibitem{AG44} T.~Alfrey~Jr. and G.~Goldfinger, J. Chem. Phys. {\bf 12}, 205 (1944).

\bibitem{FR99} G.~Fink and W.~J.~Richter, in: J.~Brandrup, E.~H.~Immergut, and E.~A.~Grulke, Editors, {\it Polymer Handbook}, 4th Edition (Wiley, New York, 1999) pp. 329-337.

\bibitem{DI48} F.~S.~Dainton and K.~J.~Ivin, Nature {\bf 162}, 705 (1948).

\bibitem{AG08} D.~Andrieux and P.~Gaspard, Proc. Natl. Acad. Sci. USA {\bf 105}, 9516 (2008).

\bibitem{AG09} D.~Andrieux and P.~Gaspard, J. Chem. Phys. {\bf 130}, 014901 (2009).

\bibitem{CT06} T.~M.~Cover and J.~A.~Thomas, {\it Elements of Information Theory} (Wiley, Hoboken, 2006).

\bibitem{J08} C.~Jarzynski, Proc. Natl. Acad. Sci. USA {\bf 105}, 9451 (2008).

\bibitem{AG13} D.~Andrieux and P.~Gaspard, EPL {\bf 103}, 30004 (2013).

\bibitem{GHJJKS09} R.~G.~Gilbert, M.~Hess, A.~D.~Jenkins, R.~G.~Jones, P.~Kratochvil, and R.~F.~T.~Stepto, Pure Appl. Chem. {\bf 81}, 351 (2009).

\bibitem{F53} P.~J.~Flory, {\it Principles of Polymer Chemistry} (Cornell University Press, Ithaca, 1953).

\bibitem{M00} C.~D.~Meyer, {\it Matrix Analysis and Applied Linear Algebra} (SIAM, Philadelphia PA, 2000).

\bibitem{G76} D.~T.~Gillespie, J. Comput. Phys. {\bf 22}, 403 (1976).

\bibitem{G77} D.~T.~Gillespie, J. Phys. Chem. {\bf 81}, 2340 (1977).

\bibitem{OCY97} K.~Osakada, J.-C.~Choi, and T.~Yamamoto,  J. Am. Chem. Soc. {\bf 119}, 12390 (1997).

\bibitem{YMIMSMMNTOYMF04} Y.~Yoshida, J.-I.~Mohri, S.-I.~Ishii, M.~Mitani, J.~Saito, S.~Matsui, H.~Makio, T.~Nakano, H.~Tanaka, M.~Onda, Y.~Yamamoto, A.~Mizuno, and T.~Fujita, J. Am. Chem. Soc. {\bf 126}, 12023 (2004).

\bibitem{ZN05} H.~Zhang and K.~Nomura,  J. Am. Chem. Soc. {\bf 127}, 9364 (2005).

\bibitem{MSSKO06} Y.~Miura, T.~Shibata, K.~Satoh, M.~Kamigaito, and Y.~Okamoto,  J. Am. Chem. Soc. {\bf 128}, 16026 (2006).

\bibitem{SMNK10} K.~Satoh, M.~Matsuda, K.~Nagai, and M.~Kamigaito,  J. Am. Chem. Soc. {\bf 132}, 10003 (2010).

\bibitem{LZFWLZ13} W.~Liu, K.~Zhang, H.~Fan, W.-J.~Wang, B.-G.~Li, and S.~Zhu, J. Polym. Sci. A: Polym. Chem. {\bf 51}, 405 (2013).

\bibitem{WLHLZW13} G.-X.~Wang, M.~Lu, Z.-H.~Hou, J.~Li, M.~Zhong, and H.~Wu, J. Polym. Sci. A: Polym. Chem. {\bf 51}, 2919 (2013).

\bibitem{DPDMJ96} A.~Duda, S.~Penczek, P.~Dubois, D.~Mecerreyes, and R.~J\'er\^ome, Macromol. Chem. Phys. {\bf 197}, 1273 (1996).

\bibitem{DK09} A.~Duda and A.~Kowalski, in: P.~Dubois, O.~Coulembier, and J.-M.~Raquez, {\it Handbook of Ring-Opening Polymerization} (Wiley-VCH, Weinheim, 2009) pp. 1-51.

\bibitem{NP13} O.~Nuyken and S.~D.~Pask, Polymers {\bf 5}, 361 (2013).

\bibitem{WSH12} S.~Whitelam, R.~Schulman, and L.~Hedges, Phys. Rev. Lett. {\bf 109}, 265506 (2012).

\bibitem{J93} K.~A.~Johnson, Annu. Rev. Biochem. {\bf 62}, 685 (1993).

\end{thebibliography}
\end{document}